\documentclass[preprint,11pt]{elsarticle}

\usepackage{amsmath,amssymb,amsthm,amsfonts}
\usepackage{mathabx}
\usepackage{mathrsfs}
\usepackage{bm}
\usepackage{graphicx}
\usepackage{float}
\usepackage{subfig}
\usepackage{color,xcolor}
\usepackage{tabularx}
\biboptions{numbers,sort&compress}
\setcitestyle{authoryear,round}
\usepackage{epsfig,psfrag}
\usepackage{tikz}
\usepackage{pgflibraryshapes}
\usepackage{graphicx,psfrag}
\usepackage{hyperref}
\usepackage{enumitem}

\usepackage{changes}
\makeatletter
\@namedef{Changes@AuthorColor}{blue}
\colorlet{Changes@Color}{blue}
\makeatother

\makeatletter
\hypersetup{colorlinks=true}
\AtBeginDocument{\@ifpackageloaded{hyperref}
	{\def\@linkcolor{magenta}
		\def\@anchorcolor{black}
		\def\@citecolor{teal}
		\def\@filecolor{cyan}
		\def\@urlcolor{magenta}
		\def\@menucolor{red}
		\def\@pagecolor{cyan}
		\begingroup
		\@makeother\`%
		\@makeother\=%
		\edef\x{%
			\edef\noexpand\x{%
				\endgroup
				\noexpand\toks@{%
					\catcode 96=\noexpand\the\catcode`\noexpand\`\relax
					\catcode 61=\noexpand\the\catcode`\noexpand\=\relax
				}%
			}%
			\noexpand\x
		}%
		\x
		\@makeother\`
		\@makeother\=
	}{}}
\makeatother

\makeatletter \oddsidemargin 0.0cm \evensidemargin 0.0cm
\newcommand{\be}{\begin{equation}}
\newcommand{\en}{\end{equation}}

\def\bm#1{\mbox{\boldmath{$#1$}}}

\numberwithin{equation}{section}
\theoremstyle{plain}

\newtheorem{theorem*}{Theorem}

\theoremstyle{definition}

\DeclareMathOperator{\sech}{sech}

\journal{ }

\begin{document}
	
\begin{frontmatter}
		
\title{ \textbf{Quantitative assessment of the infinite-domain approximation for elastic localization in finite structures and its implications for weakly nonlinear analysis}}
			
\author[mymainaddress]{Xiang Yu\corref{mycorrespondingauthor}}
\cortext[mycorrespondingauthor]{Corresponding author}
\ead{yuxiang@dgut.edu.cn}

\address[mymainaddress]{Department of Mathematics, School of Computer Science and Technology, Dongguan University of Technology, Dongguan, 523808, China}

 \begin{abstract}
 A common approach to studying elastic localization is to pose the
 problem on an infinite domain, avoiding explicit treatment of the end
 conditions. For a finite structure, this approach raises a natural
 question: how long must the structure be for the infinite-domain
 approximation to become accurate, and how large is the resulting error?
 To address this question, we develop a constructive method that gives
 a length threshold beyond which the approximation error is small and
 explicitly bounded. For a prototype model arising from twisted-rod
 buckling, the threshold and error bounds are obtained analytically. The
 threshold varies with loading, showing that the accuracy of the
 infinite-domain approximation depends on both structural length and
 loading state. The load-dependent threshold also explains why the
 finite-domain weakly nonlinear expansion is accurate only very close
 to bifurcation, while the infinite-domain expansion can remain accurate
 over a broader loading range. The method is then applied to localized bulging in cylindrical membrane
 tubes, for which the model-dependent estimates are evaluated numerically.
 Fully nonlinear computations support the predicted length threshold and
 the weakly nonlinear validity regimes. The full rod equation includes
 an end correction omitted in the prototype model. The constructive
 method still gives the threshold and error bound in closed form.
 Together, these results characterize the applicability of the
 infinite-domain approximation and weakly nonlinear descriptions to
 elastic localization in finite structures.
 \end{abstract}

\end{frontmatter}

\section{Introduction}

Localization occurs in a wide range of solids and structures,
including necking in solids
\citep{hill1952discontinuous,hutchinson1983neck,audoly2016analysis,
	springhetti2024necking}, bulging in membrane tubes
\citep{haughton1979bifurcation,kyriakides1990inflation,
	kyriakides1991initiation,fu2008post,lestringant2018diffuse,
	yu2023one,wang2019experimental}, localized buckling in twisted rods
\citep{coyne1990analysis,thompson1999helix,van2000helical}, and
elasto-capillary necking in soft cylinders
\citep{taffetani2015beading,lestringant2020one,fu2021necking,
	xuan2017plateau,li2022surface,giudici2020ballooning}. In these problems, an initially
homogeneous deformation loses stability and develops into an
inhomogeneous state in which deformation becomes concentrated within
part of the structure. The associated strain concentration can strongly
affect the load response and may lead to material or structural failure
\citep{bordignon2015strain,briffod2026understanding}. Understanding how
localized states emerge and evolve along the post-bifurcation path is
therefore an important problem in nonlinear elasticity and structural
mechanics.

A common approach to studying localization is to pose the problem on an infinite
domain, replacing the physical end conditions by decay to a homogeneous
state at infinity
\citep{coyne1990analysis,van2000helical,fu2008post,
	pearce2010characterization,wang2021necking}. This replacement simplifies the analysis but changes the boundary-value problem. In general, the localized solution selected by the finite-domain end conditions is distinct from the infinite-domain solution restricted to the same interval. For twisted rods, \cite{neukirch2002writhing} showed that finite length can change when the rod loses stability and introduce instabilities absent in the infinite-rod approximation..
Infinite-domain approximations can
nevertheless be accurate for structures of moderate length
\citep{wang2021necking}. For a finite structure, a natural question therefore arises: how long must the structure be for the infinite-domain approximation to become accurate, and how large is the error beyond that length?

The relation between finite- and infinite-domain localized solutions has
long been studied in dynamical systems theory. In dynamical-systems
terms, finite- and infinite-domain localized solutions correspond,
respectively, to periodic and homoclinic orbits. The approach of the periodic orbit to the homoclinic orbit as its
period tends to infinity is well known in dynamical systems theory
\citep{guckenheimer1983nonlinear,champneys1998homoclinic}. Under
appropriate non-degeneracy conditions, results from dynamical
systems theory establish the
existence of such periodic-orbit families and their convergence to the
homoclinic orbit as the period tends to infinity
\citep{devaney1977blue,vanderbauwhede1992homoclinic,
	vanderbauwhede1998branching}; exponential estimates for this convergence
are also available \citep{vanderbauwhede1992homoclinic}. These results provide the asymptotic basis for the infinite-domain
approximation. In these abstract formulations, however, the
system-dependent constants in the estimates and the period thresholds
beyond which the estimates apply generally remain undetermined.

To address this question, we develop a constructive method for
quantitatively assessing the infinite-domain approximation. The method yields an explicit length criterion and an associated bound on the approximation error. We first illustrate the method using a prototype localization model,
whose simple form allows the relevant constants to be determined
analytically. For this model, the length criterion takes the explicit form
\begin{equation}\label{eq:intro-length-criterion}
L\geq L_*(\lambda).
\end{equation}
Here \(L\) is the structural length, \(\lambda\) is the loading
parameter, and \(L_*(\lambda)\) is the explicitly determined length
threshold, which varies with $\lambda$. When this criterion is satisfied, the infinite-domain approximation
achieves the prescribed accuracy, with an explicit bound on the
approximation error. The length
criterion therefore shows that whether the infinite-domain
approximation is accurate depends on both structural length and
loading state.

\begin{figure}[h!]
	\centering
	\includegraphics[width=0.9\linewidth]{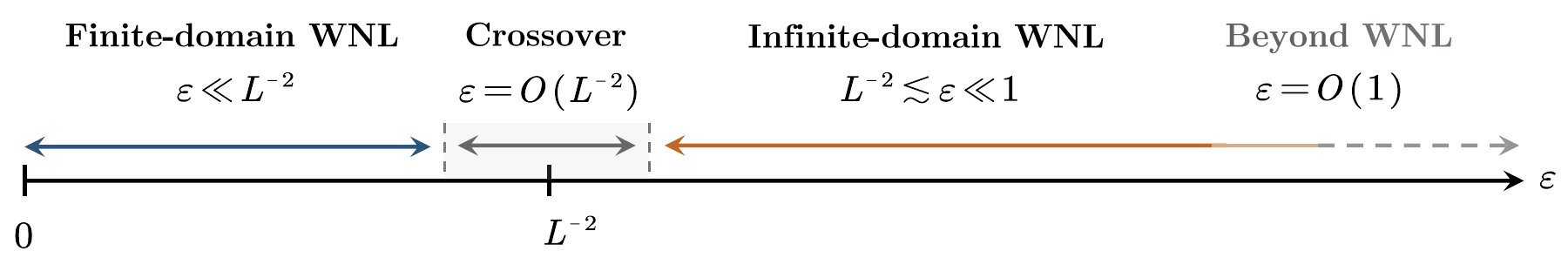}
	\caption{Schematic illustration of the validity regimes of the
		finite- and infinite-domain weakly nonlinear (WNL) approximations
	to a finite-domain localized solution.}
	\label{fig:wnl}
\end{figure}

The load-dependent length threshold also clarifies a contrast in the
validity of weakly nonlinear approximations noted in previous studies
of elastic localization
\citep{fu2008post,lestringant2018diffuse,lestringant2020one,wang2021necking},
namely that the finite-domain expansion is accurate only very close to
bifurcation, whereas the infinite-domain expansion can remain accurate
over a broader loading range. With \(\varepsilon\) measuring the distance of the loading parameter from its bifurcation value, the length
criterion identifies \(\varepsilon=O(L^{-2})\) as the crossover regime beyond which the fully nonlinear
finite-domain solution is well approximated by its infinite-domain
counterpart. Very close to bifurcation, the finite-domain expansion provides
the appropriate local description. Beyond the crossover regime, the closeness of
the two fully nonlinear solutions makes the infinite-domain expansion
more appropriate. As \(L\) increases, the crossover moves closer to
bifurcation, narrowing the validity range of the finite-domain
expansion. Fig.~\ref{fig:wnl} summarizes these regimes schematically.

We then apply the constructive method to two concrete localization problems in
mechanics, namely localized bulging in cylindrical membrane tubes and
localized buckling in twisted rods. For the membrane tube, numerical evaluation of the
model-dependent estimates gives the length threshold and an upper bound
on the approximation error. Comparisons with the fully nonlinear
solutions support the predicted crossover and show how the validity
range of the finite-domain weakly nonlinear expansion narrows as the
tube length increases. For the twisted rod, we show how the prototypical
model follows from the rod equilibrium equations and which end correction
it omits. With this correction included, the constructive method yields
the length threshold and an exponential error bound analytically.

We further apply the constructive method to the fully nonlinear problem
of localized bulging in cylindrical membrane tubes. For this problem, the model-dependent estimates entering the
constructive method are obtained numerically. These estimates yield the 
length threshold and an upper bound on the approximation
error. Comparisons with the fully nonlinear solutions support the predicted
crossover and show how the validity range of the finite-domain weakly
nonlinear expansion narrows as the tube length increases. For the twisted rods, we show how the prototypical model arises from
localized buckling. The model omits an end correction present in the
finite-rod equation. For the full rod problem, the constructive method still
yields the length threshold and an exponential error bound analytically.

The remainder of the paper is organized as follows. Section~\ref{sec:prob} formulates the finite- and infinite-domain prototypical problems. Section~\ref{sec:length} develops the constructive method and derives the length criterion and error bounds. Section~\ref{sec:weakly} examines the validity ranges of  weakly nonlinear analysis. Section~\ref{sec:application} applies the method to membrane-tube bulging, and Section~\ref{sec:connection} connects the prototypical model to twisted-rod buckling and applies the method to the full
rod problem. Section~\ref{sec:con} concludes the paper.

\section{A prototype model for the infinite-domain approximation}
\label{sec:prob}

We first use a prototype localization model to formulate the finite- and infinite-domain problems needed to assess the infinite-domain approximation. The model arises from a simplified description of localized buckling in a twisted
elastic rod \citep{coyne1990analysis,van2000helical}. Its main role here is to illustrate the method developed in Section~\ref{sec:length}, with each step worked out explicitly.

On the half-domain,  the finite-domain  problem is
\begin{align}
&y''(x)
=\alpha(\lambda)y(x)+\beta(\lambda)y^2(x),\qquad x\in [0,L],
\label{eq:orig}\\
&y'(0)=0,
\qquad
y'(L)=0.
\label{eq:original-bc}
\end{align}
Here \(x\) denotes the axial coordinate, \(L\) is the structural
half-length, referred to simply as the length, \(y(x)\) is a scalar measure of the deformation that is assumed to be monotonic on \([0,L]\),  $\lambda$ is a loading parameter, and the coefficients \(\alpha\) and \(\beta\)
depend on $\lambda$.  The condition \(y'(0)=0\) follows from symmetry about the
localization center, whereas \(y'(L)=0\) is the physical end condition
on the finite structure. All lengths are nondimensionalized by a fixed transverse length scale.

The corresponding infinite-domain problem retains the same
equilibrium equation and symmetry condition at the localization center,
while replacing the boundary condition at the finite end by the decay
condition at infinity:
\begin{align}
&y_\infty''(x)
=\alpha(\lambda)y_\infty(x)
+\beta(\lambda)y_\infty^2(x),
\qquad x\in[0,\infty),
\label{eq:orig-infinite}\\
&y_\infty'(0)=0,
\qquad
\lim_{x\to\infty}y_\infty(x)=0,
\label{eq:orig-infinite-bc}
\end{align}
where \(y_\infty(x)\) denotes the infinite-domain solution used to approximate \(y(x)\) at the same loading state.

We assume \(\alpha(\lambda)>0\) and $\beta(\lambda)\neq 0$, which supports
the localized solution branch. To facilitate analysis, we introduce the normalized variables
\begin{equation}\label{eq:normalization}
s=\sqrt{\alpha(\lambda)} x,
\qquad
u(s)=-\frac{\beta(\lambda)}{\alpha(\lambda)}y(x),
\qquad
u_\infty(s)
=-\frac{\beta(\lambda)}{\alpha(\lambda)}y_\infty(x).
\end{equation}
 The finite-domain problem
\eqref{eq:orig}--\eqref{eq:original-bc} then becomes
\begin{align}
&u''(s)=u(s)-u^2(s),
\qquad s\in[0,\ell],
\label{eq:model_scaled}\\
&u'(0)=0,
\qquad
u'(\ell)=0,
\label{eq:bc}
\end{align}
where $\ell=\sqrt{\alpha(\lambda)}L$ is the normalized half-length. The corresponding infinite-domain problem takes the form
\begin{align}
u_\infty''(s)
&=u_\infty(s)-u_\infty^2(s),
\qquad s\in[0,\infty),
\label{eq:model-infinite}\\
u_\infty'(0)&=0,
\qquad
\lim_{s\to\infty}u_\infty(s)=0.
\label{eq:bc-infinite}
\end{align}
The normalized equilibrium equations are independent of \(\lambda\). The load enters through the half-length \(\ell\) and the transformation back to the original variables. The localized solution of the infinite-domain problem is given explicitly by
\begin{equation}\label{eq:yinf}
u_\infty(s)
=\frac{3}{2}\sech^2\Big(\frac{s}{2}\Big).
\end{equation}

\section{A constructive method for quantifying the infinite-domain approximation}\label{sec:length}

This section develops a constructive method for determining when the infinite-domain approximation is accurate for a finite structure and quantifying the approximation error.

\subsection{Phase-space structure and the length integral}
\label{sec:phase-space}

\begin{figure}[h!]
	\centering
	\subfloat[]{
		\includegraphics[width=0.43\textwidth]{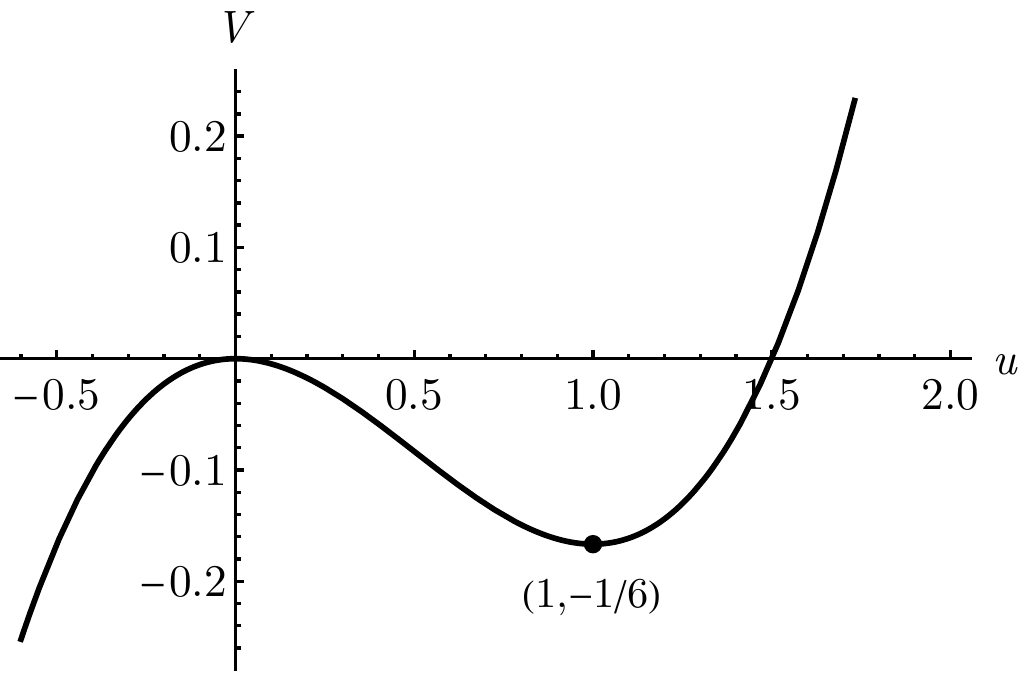}
	}
	\qquad\qquad
	\subfloat[]{
		\includegraphics[width=0.43\textwidth]{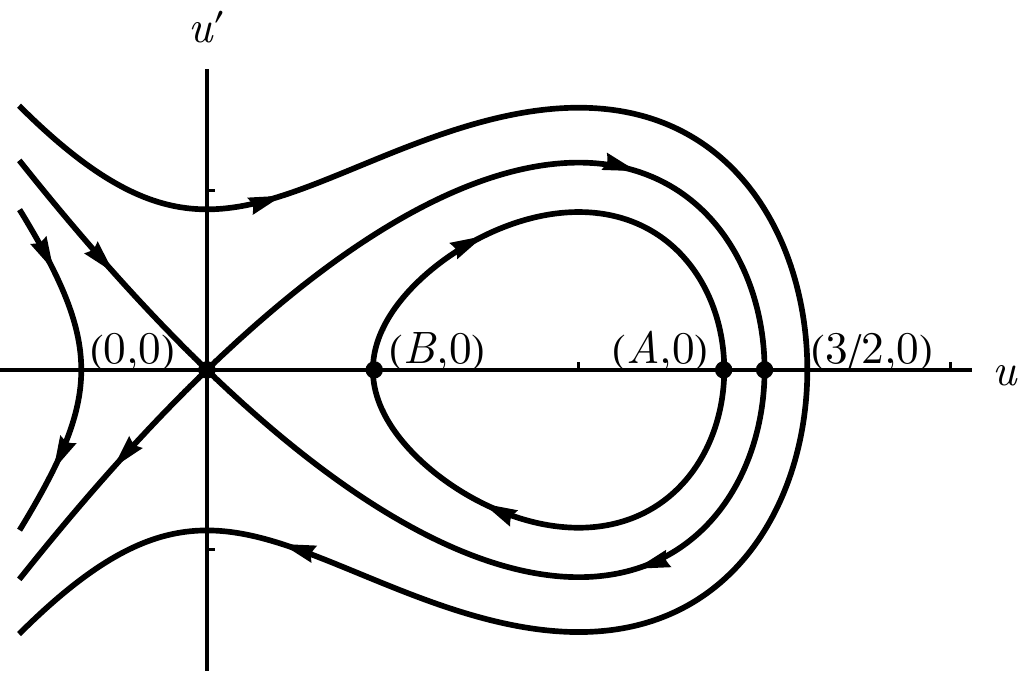}
	}
	\caption{
	(a) The potential
	\(V(u)=-\frac12u^2+\frac13u^3\).
	(b) Phase portrait of \eqref{eq:model_scaled}. The arrows indicate the
	direction of increasing \(s\). 
	}
	\label{fig:phase-plane}
\end{figure}

We begin with a brief phase-space description of the two solutions being compared. Multiplying \eqref{eq:model_scaled} by \(u'\) gives the
first integral
\begin{equation}\label{eq:first-integral}
\frac12 u'^2+V(u)=H,
\qquad
V(u)=-\frac12u^2+\frac13u^3,
\end{equation}
where \(H\) is a constant of integration. The level sets of \eqref{eq:first-integral} determine the phase-space
orbits. The equilibria \((0,0)\) and \((1,0)\) are a saddle and a center,
respectively. For \(-1/6<H<0\), the bounded level sets are periodic
orbits surrounding \((1,0)\), while \(H=0\) contains the homoclinic orbit.
The potential and representative orbits are shown in
Fig.~\ref{fig:phase-plane}.

We restrict attention to the finite-domain solution for which
\(u(s)\) decreases monotonically on \(s\in[0,\ell]\). Set $A=u(0)$ and $B=u(\ell)$.
In the phase plane, \(u(s)\) traces the lower half of a periodic orbit
from \((A,0)\) to \((B,0)\), whereas \(u_\infty(s)\) traces the lower
branch of the homoclinic orbit from \((3/2,0)\) towards \((0,0)\). Hence \(0<B<1<A<3/2\).
Since \(u'(0)=u'(\ell)=0\), evaluating
\eqref{eq:first-integral} at the two endpoints gives
\begin{equation}\label{eq:turning-point-relation}
V(A)=V(B)=H.
\end{equation}
Solving for the root satisfying \(1<A<3/2\) gives the turning-point relation
\begin{equation}\label{eq:A-of-B}
A
=\frac{3-2B+\sqrt{9+12B-12B^2}}{4}.
\end{equation}

Because \(u'<0\) on \(s\in[0,\ell]\), using \(H=V(B)\) in
\eqref{eq:first-integral} and separating variables gives
\begin{equation}\label{eq:l_integral}
\ell
=\int_B^{A}
\frac{du}{\sqrt{2(H-V(u))}}
=\int_B^{A}
\frac{du}
{\sqrt{u^2-B^2-\frac23(u^3-B^3)}}.
\end{equation}
The integral determines \(\ell\) from \(B\). As \(B\to1^-\),
\(\ell\to\pi\), so the finite-domain localization branch exists only for
\(\ell>\pi\). As \(B\to0^+\), one has \(A\to3/2\), \(H\to0^-\), and
\(\ell\to\infty\), while the periodic orbit approaches the homoclinic orbit
corresponding to \(u_\infty\).

\subsection{Quantitative estimate of the length integral}
\label{sec:length-estimate}

To quantify the divergence of the length integral as \(B\to0^+\), we set
\(\delta=\sqrt{B}\) and split \eqref{eq:l_integral}  as
\(\ell=I_1+I_2\), where
\begin{align}\label{eq:I1I2-main}
I_1&=
\int_B^\delta
\frac{du}
{\sqrt{u^2-B^2-\frac23(u^3-B^3)}},
\qquad
I_2=
\int_\delta^A
\frac{du}
{\sqrt{u^2-B^2-\frac23(u^3-B^3)}}.
\end{align}

The first integral is governed by the quadratic terms near the lower
turning point \(u=B\), whereas the second approaches the homoclinic
integral. As \(B\to0^+\), we obtain
\begin{align}
I_1
&=\int_B^\delta\frac{du}{\sqrt{u^2-B^2}}+o(1)
=\ln\frac{2}{\delta}+o(1),\\
I_2
&=\int_\delta^{3/2}\frac{du}{\sqrt{u^2-\frac{2}{3}u^3}}+o(1)
=\ln\frac{6}{\delta}+o(1).
\end{align}
The derivation of these estimates for \(I_1\) and \(I_2\) is given
in \ref{app:length-evaluation}.  Adding the two expressions and using \(\delta^2=B\),
we obtain the length relation
\begin{align}\label{eq:length-amplitude-explicit}
\ell=-\ln B+\ln 12+R(B),
\end{align}
where \(R(B)\to0\) as \(B\to0^+\). For the range of endpoint values
used below, \ref{app:length-evaluation} gives the explicit bound
\begin{align}\label{eq:length-remainder-bound}
0<R(B)\leq 0.012,
\qquad 0<B\leq0.03.
\end{align}

\subsection{Explicit length criterion and error bounds}\label{sec:criterion}

The finite-domain solution converges to the infinite-domain solution
as the normalized length $\ell$ tends to infinity. To determine whether
the two solutions are close at a finite length, we need a quantitative
criterion. We require their maximum difference over the interval to be at most one
percent of the peak value of the finite-domain solution, consistent
with relative errors of about one percent for wave approximations
\citep{boyd1986solitons,parker1995exact}.

At the endpoint, this criterion requires
\begin{align}\label{eq:endpoint-criterion}
|B-u_\infty(\ell)|\leq0.01A.
\end{align}
By \eqref{eq:length-amplitude-explicit} and \eqref{eq:yinf}, we can
write
\begin{align}
u_\infty(\ell)=\frac{B}{2} \frac{e^{-R(B)}}{(1+e^{-\ell})^2}
\end{align}
Since \(R(B)>0\), we have \(u_\infty(\ell)\leq B/2\), so
\eqref{eq:endpoint-criterion} requires \(B\leq0.02A<0.03\).
For \(B\) in this range, \(R(B)\leq0.012\) by
\eqref{eq:length-remainder-bound}, and \(e^{-\ell}\leq B/12\).
Substituting these bounds into the expression for \(u_\infty(\ell)\)
gives
\begin{align}
0.50B\leq B-u_\infty(\ell)\leq0.51B.
\end{align}

To impose \eqref{eq:endpoint-criterion}, we allow a small margin to
account for the use of inequalities. The estimates used throughout the
argument are reasonably  tight, so we increase the bound \(0.51B\) by ten percent
and require \(1.1\times 0.51B\leq 0.01A\). Since
\(A\geq3/2-2B^2/3\geq1.4994\) for \(B\leq0.03\), this requirement
is met whenever
\begin{align}\label{eq:Bineq}
B\leq B_0,
\qquad
B_0=\frac{0.01\times1.4994}{1.1\times 0.51}=\frac{147}{5500}.
\end{align}
To express this condition in terms of length, we combine
\eqref{eq:length-amplitude-explicit} with the remainder bound
\eqref{eq:length-remainder-bound} to obtain
\begin{align}\label{eq:BB}
B=12e^{-\ell+R(B)}\leq12.2e^{-\ell}.
\end{align}
Requiring \(12.2e^{-\ell}\leq B_0\) then ensures that the endpoint
criterion \eqref{eq:endpoint-criterion} holds for all
\begin{align}\label{eq:ell-star}
\ell\geq\ell_*,
\qquad
\ell_*=\ln\frac{12.2}{B_0}\approx6.12.
\end{align}

For lengths at or above this threshold, the relations \eqref{eq:A-of-B} and \eqref{eq:BB} give
the error estimate at the center:
\begin{align}
0<\frac32-A
\leq \frac23 B^2
\leq \frac23(12.2)^2 e^{-2\ell}.
\end{align}
Let \(w(s)=u(s)-u_\infty(s)\) denote the approximation error.  Subtracting the
governing equation for $u_\infty$ from that for $u$ gives
\begin{align}\label{eq:erreq}
w''
&=(1-2u_\infty)w-w^2,
\qquad
w(0)=A-\frac32,\qquad w'(0)=0.
\end{align}
The linear part of \eqref{eq:erreq} governs the propagation of the
initial error \(w(0)\) from the center toward the endpoint, while a
bootstrap argument \citep{tao2006nonlinear} controls the nonlinear
correction. As shown in \ref{app:uniform-error}, the resulting
bound on the maximum error is
\begin{align}\label{eq:uniform-normalized-error}
6e^{-\ell}
\leq \max_{0\leq s\leq\ell}|u(s)-u_\infty(s)|
\leq 6.6e^{-\ell},
\qquad \ell\geq\ell_*,
\end{align}
where the lower bound follows from the error at the endpoint. Since \(A\geq1.4994\), this estimate guarantees a maximum relative
error of at most \(6.6e^{-\ell_*}/1.4994\approx0.964\%<1\%\)
for every \(\ell\geq\ell_*\). A direct numerical comparison of the
two solutions gives a maximum relative error of \(1\%\) at
\(\ell\approx6.02\), close to the predicted threshold \(\ell_*\approx6.12\).

Returning to the original variables, we define the load-dependent
length threshold
\begin{align}\label{eq:physical-length-threshold}
L_*(\lambda)
:=
\frac{\ell_*}{\sqrt{\alpha(\lambda)}}.
\end{align}
The condition \(\ell\geq\ell_*\) is then equivalent to the length
criterion \(L\geq L_*(\lambda)\). Whenever this criterion is satisfied,
\eqref{eq:uniform-normalized-error} gives
\begin{equation}\label{eq:uniform-physical-error}
6\frac{\alpha(\lambda)}{|\beta(\lambda)|}
e^{-\sqrt{\alpha(\lambda)}L} \leq \sup_{ x\in[0,L]}
|y(x)-y_\infty(x)|
\leq
6.6 \frac{\alpha(\lambda)}{|\beta(\lambda)|}
e^{-\sqrt{\alpha(\lambda)}L},
\qquad
L\geq L_*(\lambda).
\end{equation}
Thus, for \(L\geq L_*(\lambda)\), the maximum difference over the
entire interval is at most one percent of \(\lvert y(0)\rvert\) and
decays exponentially as the length increases. Since \(L_*(\lambda)\) depends on the loading parameter, whether the
localized solution in a fixed-length structure can be well approximated
by its infinite-domain counterpart depends on the loading state.

The exponential decay in \eqref{eq:uniform-physical-error} is consistent with large-period homoclinic estimates from dynamical systems theory \citep{vanderbauwhede1992homoclinic}; an exponential estimate for the prototype equation is also given by \citet{pelinovsky2012enstrophy}. The analysis above additionally specifies the length threshold and error prefactors for the prototype model.  The finite-domain solution also admits a Jacobi elliptic-function representation. The relation between the domain length and the elliptic parameter involves a complete elliptic integral and is an alternative parametrization of the length--endpoint relation analyzed above. The constructive method makes the length threshold and error prefactors
explicit without relying on this representation.

The constructive method therefore turns the requirement that a
structure be ``sufficiently long'' into the quantitative, load-dependent
criterion \(L\geq L_*(\lambda)\), and provides explicit bounds on the
approximation error once the criterion is satisfied.

\section{Validity of weakly nonlinear analysis}\label{sec:weakly}

Having quantified the infinite-domain approximation, we now use the length criterion to examine the validity ranges of the finite- and infinite-domain weakly nonlinear descriptions. The analysis identifies the crossover between the finite- and infinite-domain weakly nonlinear descriptions and determines how this crossover depends on structural length.

\subsection{Finite-domain weakly nonlinear analysis}\label{sec:ph}

For $\alpha(\lambda)>0$, the finite-domain problem \eqref{eq:orig}--\eqref{eq:original-bc} has a saddle equilibrium at $y=0$ and a center equilibrium at $y=-\alpha(\lambda)/\beta(\lambda)$. The finite-domain branch bifurcates from the center, where small periodic orbits around it can satisfy the Neumann boundary conditions. We therefore set $z(x)=y(x)+\alpha(\lambda)/\beta(\lambda)$, which satisfies
\begin{align}
&z''(x)=-\alpha(\lambda)z(x)+\beta(\lambda)z(x)^2,\qquad x\in[0,L],\\
&z'(0)=0,\qquad z'(L)=0.
\end{align}
Linearizing about $z=0$ gives $z''=-\alpha(\lambda)z$. The first nontrivial mode is $z=A\cos(\pi x/L)$, where $A$ denotes the modal amplitude. The corresponding bifurcation condition is
\begin{equation}\label{eq:bif}
\alpha(\lambda_{\mathrm{cr}})=\frac{\pi^2}{L^2},
\end{equation}
where $\lambda_{\mathrm{cr}}$ denotes the critical loading parameter. For notational convenience, we use the shorthand \(\alpha'_{\mathrm{cr}}=\alpha'(\lambda_{\mathrm{cr}})\) and \(\beta_{\mathrm{cr}}=\beta(\lambda_{\mathrm{cr}})\).

Near the bifurcation point, writing $\lambda
=\lambda_{\mathrm{cr}}
+\varepsilon\lambda_0$, 
where $\varepsilon \ll 1$ and $\lambda_0=O(1)$, we seek an expansion of the form
\begin{equation}\label{eq:finite-expansion}
z(x)
=
\varepsilon^{1/2}A\cos \frac{\pi x}{L}
+\varepsilon z_2(x)
+\varepsilon^{3/2}z_3(x)
+\cdots .
\end{equation}
Substitution into the governing equation gives, at  $O(\varepsilon)$,
\begin{equation}\label{eq:y2}
z_2(x)
=
\frac{\beta_{\mathrm{cr}}A^2L^2}{2\pi^2}
\Big(
1-\frac{1}{3}\cos\frac{2\pi x}{L}
\Big).
\end{equation}
At $O(\varepsilon^{3/2})$, the solvability condition is
\begin{equation}\label{eq:solvability-finite}
\lambda_0\alpha'_{\mathrm{cr}}A
-
\frac{5\beta_{\mathrm{cr}}^2L^2}{6\pi^2}A^3
=
0.
\end{equation}
For the branch continuing toward the localized solution, we choose $\beta_{\mathrm{cr}} A<0$.  Substituting the
corresponding root of \eqref{eq:solvability-finite}, together with
\eqref{eq:y2}, into \eqref{eq:finite-expansion} gives the two-term
finite-domain weakly nonlinear approximation
\begin{align}\label{eq:finite-wnl}
z(x)
=-\sqrt{\varepsilon}\,\frac{\pi}{\beta_\text{cr}L}
\sqrt{
	\frac{6\lambda_0\alpha'_{\mathrm{cr}}}
	{5}
}
\cos \frac{\pi x}{L} +
\varepsilon
\frac{\lambda_0\alpha'_{\mathrm{cr}}}
{5\beta_{\mathrm{cr}}}
\Big(3-\cos \frac{2\pi x}{L}\Big)
+O(\varepsilon^{3/2}).
\end{align}

The two terms in \eqref{eq:finite-wnl} have characteristic magnitudes
\(O(\sqrt{\varepsilon}/L)\) and \(O(\varepsilon)\), respectively.
Maintaining their asymptotic ordering therefore requires  $
\varepsilon
\ll
{\sqrt{\varepsilon}}/{L}$,
or equivalently
\begin{equation}\label{eq:wnl_validity}
\varepsilon\ll\frac{1}{L^2}.
\end{equation}
Thus, the finite-domain expansion is valid as \(\varepsilon\to0\) for any
fixed \(L\), but not uniformly in \(L\), since its validity range narrows as
the structural length increases.

\subsection{Infinite-domain weakly nonlinear analysis}\label{sec:inf}

We next consider the infinite-domain problem \eqref{eq:orig-infinite}--\eqref{eq:orig-infinite-bc}. The bifurcation condition is
\begin{equation}
\alpha(\lambda^\infty_{\mathrm{cr}})=0,
\end{equation}
where $\lambda_{\mathrm{cr}}^\infty$ denotes the infinite-domain critical loading parameter at which the linearized equation changes character. This may be viewed as the limit of the finite-domain bifurcation
condition \eqref{eq:bif} as \(L\to\infty\).

Near the critical state, we write $\lambda
=
\lambda^\infty_{\mathrm{cr}}
+\varepsilon\lambda_0$,
where $\varepsilon \ll 1$ and $\lambda_0=O(1)$. Balancing the linear,
nonlinear, and second-derivative terms shows that
\(y_\infty=O(\varepsilon)\) and varies on the spatial scale
\(O(\varepsilon^{-1/2})\). We therefore introduce
\begin{equation}
 y_\infty(x)=\varepsilon y_1(S)+\cdots,\qquad S=\sqrt{\varepsilon}x.
\end{equation}
At leading order, the governing equation reduces to
\begin{equation}
y_1''(S)
=
\lambda_0\alpha'(\lambda_\text{cr}^\infty)y_1(S)
+\beta(\lambda_\text{cr}^\infty)y_1(S)^2,
\end{equation}
whose localized solution gives
\begin{equation}\label{eq:infinite_wnl}
y_\infty(x)
=
-\varepsilon
\frac{3\lambda_0\alpha'(\lambda_\text{cr}^\infty)}
{2\beta(\lambda_\text{cr}^\infty)}
\sech^2
\Big(
\frac{\sqrt{\varepsilon\lambda_0\alpha'(\lambda_\text{cr}^\infty)}}{2}x
\Big)
+O(\varepsilon^2).
\end{equation}
As an approximation to the infinite-domain solution, the
infinite-domain weakly nonlinear approximation requires only
\begin{equation}\label{eq:infinite_wnl_validity}
\varepsilon\ll1.
\end{equation}

\subsection{Validity ranges of the weakly nonlinear approximations}

For a structure of fixed length \(L\), the criterion
\(L\geq L_*(\lambda)\) guarantees that the finite- and infinite-domain
solutions are close throughout the interval. The crossover loading
parameter is determined by
\begin{equation}\label{eq:crossover-alpha}
L
=
\frac{\ell_*}{\sqrt{\alpha(\lambda)}}\quad\iff \quad \alpha(\lambda)
=
\frac{\ell_*^2}{L^2}.
\end{equation}
Combining \eqref{eq:bif} and \eqref{eq:crossover-alpha} gives, at the crossover, $
\alpha(\lambda)-\alpha(\lambda_{\mathrm{cr}})
={(\ell_*^2-\pi^2)}/{L^2}$.
Assuming $\alpha'$ remains bounded away from zero near $\lambda_{\mathrm{cr}}$, we have $\lambda-\lambda_{\mathrm{cr}}=O(L^{-2})$. Since $\lambda-\lambda_{\mathrm{cr}}=\varepsilon\lambda_0$, it follows that
\begin{equation}\label{eq}
\varepsilon=O(L^{-2}).
\end{equation}
This $O(L^{-2})$ crossover scale is measured in the loading parameter $\lambda$; its relation to the physical load depends on the specific model.

The finite-domain expansion requires $\varepsilon\ll L^{-2}$, whereas the crossover occurs at $\varepsilon=O(L^{-2})$. At the crossover, the finite-domain localized solution is close to its infinite-domain counterpart and is no longer well represented by the finite-domain critical mode. This explains the loss of validity of the finite-domain weakly nonlinear approximation.

The loss of validity of the finite-domain expansion does not mark the end of a weakly nonlinear description. Beyond the crossover, the finite-domain solution is close to its infinite-domain counterpart, so the infinite-domain weakly nonlinear expansion can be used instead. This approximation involves two errors: the difference between the finite- and infinite-domain solutions, controlled by the length criterion, and the truncation error of the infinite-domain expansion, which is asymptotically small for $\varepsilon\ll1$. The infinite-domain expansion therefore describes the finite-domain solution over the broader regime
\begin{equation}\label{eq:infinite_wnl_regime}
\frac{1}{L^2}\lesssim\varepsilon\ll1,
\end{equation}
where \(\lesssim\) denotes an inequality up to a multiplicative constant. 

 The same \(O(L^{-2})\) scaling appears in \citet{van2000helical}, but
with a different meaning. Their scaling concerns a homoclinic solution
restricted to a finite interval and the end-shortening and end rotation
calculated from it. The length criterion identifies the loading scale
at which the appropriate weakly nonlinear description changes from the
finite-domain expansion to the infinite-domain expansion.
Fig.~\ref{fig:wnl} summarizes the validity ranges of the two expansions.

\section{Application to localized bulging in membrane tubes}\label{sec:application}

We now apply the constructive method to localized bulging in cylindrical
membrane tubes \citep{fu2008post,lestringant2018diffuse}. We adopt the standard notation of the membrane-tube literature, so some symbols differ from those used in the preceding sections.

\subsection{Problem formulation and governing equations}

We consider an incompressible isotropic cylindrical membrane tube with
undeformed thickness \(H\), radius \(R\), and length \(2L\), undergoing an
axisymmetric deformation under internal pressure and an applied axial force. Assuming symmetry about the midpoint $Z=0$, we restrict attention to the half-domain $[0,L]$.

An axisymmetric deformation is described by
\begin{align}
r = r(Z), \qquad \theta = \Theta, \qquad z = z(Z),
\end{align}
where $(R,\Theta,Z)$ and $(r,\theta,z)$ denote the reference and current cylindrical coordinates, respectively. 
The principal stretches are
\begin{align}
\lambda_1 = \frac{r}{R}, \qquad 
\lambda_2 = \sqrt{r'^2 + z'^2}, \qquad 
\lambda_3 = \lambda_1^{-1} \lambda_2^{-1},
\end{align}
where a prime denotes differentiation with respect to $Z$, and $\lambda_3$ follows from incompressibility.

The total potential energy of the membrane, scaled by $2\pi R H$, is
\begin{align}\label{eq:memtp}
\mathcal{E}_{\mathrm{mem}}[r,z] 
= \int_0^L \Big(
w(\lambda_1,\lambda_2) 
- \frac{P}{2} \frac{r^2}{R} z'
- N z'
\Big) \, dZ,
\end{align}
where \(w(\lambda_1,\lambda_2)\) is the strain-energy function, $P$ is the internal pressure scaled by $H$, and $N$ is the applied axial force.  The associated Euler--Lagrange equations are
\begin{align}
&\frac{w_1}{R} 
- \Big( w_2 \frac{r'}{\lambda_2} \Big)' 
- P \frac{r}{R} z' = 0, \label{eq:mem1} \\
&w_2 \frac{z'}{\lambda_2} 
- \frac{P}{2} \frac{r^2}{R} 
- N = 0, \label{eq:mem2}
\end{align}
where $w_i = \partial w / \partial \lambda_i$ for $i=1,2$. 
These equations are supplemented by the boundary conditions
\begin{align}\label{eq:membc}
r'(0)=0, \qquad r'(L)=0,
\end{align}
where the first condition follows from symmetry and the second is the natural
boundary condition. Since the energy density has no explicit dependence on
\(Z\), the equilibrium equations admit the first integral \citep{fu2008post}
\begin{align}\label{eq:fir}
w - \lambda_2 w_2 = \text{constant}.
\end{align}

Eq.~\eqref{eq:mem2} is algebraic in \(z'\). Under the standard strong
ellipticity condition on the strain-energy function, it determines \(z'\)
implicitly as a function of \(r\) and \(r'\). Substitution into
\eqref{eq:mem1} then yields a single autonomous second-order equation 
\begin{align}\label{eq:rf}
r''=f(r,r'),
\end{align}
where \(f\) is determined by the strain-energy function and the loading
parameters. Eq. \eqref{eq:rf}, together with the boundary conditions \eqref{eq:membc}, defines the fully nonlinear finite-domain boundary-value
problem for localized bulging. The corresponding infinite-domain problem is posed on \(Z\in[0,\infty)\), with \(r'(0)=0\) and \(r(Z)\) approaching a constant as \(Z\to\infty\).

\subsection{Homogeneous deformation}

We first analyze homogeneous solutions of the form
\begin{align}
r = a R, \qquad z = \lambda Z,
\end{align}
where $a$ and $\lambda$ are constants representing the azimuthal and axial stretches, respectively. Hereafter, all lengths are normalized by the undeformed
radius \(R\), so that \(R=1\) and the homogeneous radius is \(r=a\).

The principal stretches  are then $\lambda_1 = a$ and $\lambda_2 = \lambda$.
Substitution into \eqref{eq:mem1} and \eqref{eq:mem2} gives
\begin{align}\label{eq:hom}
P = \frac{w_1(a,\lambda)}{a \lambda}, \qquad 
N = w_2(a,\lambda) - \frac{a  w_1(a,\lambda)}{2\lambda}.
\end{align}

For the computations below, we adopt the Gent material model \citep{gent1996new}, with the strain-energy function
\begin{align}
w(\lambda_1,\lambda_2) 
= -\frac{\mu J_m}{2}\ln\!\left(1 - \frac{\lambda_1^2 + \lambda_2^2 + \lambda_1^{-2}\lambda_2^{-2} - 3}{J_m}\right),
\end{align}
where $\mu$ is the shear modulus and $J_m$ is the locking parameter. Following \cite{fu2008post}, we set  
\begin{align}\label{eq:par}
\mu = 1, \qquad J_m = 97.2,
\end{align}
with \(\mu=1\) setting the stress scale.

\begin{figure}[h!]
	\centering
	\subfloat[]{\includegraphics[width=0.43\textwidth]{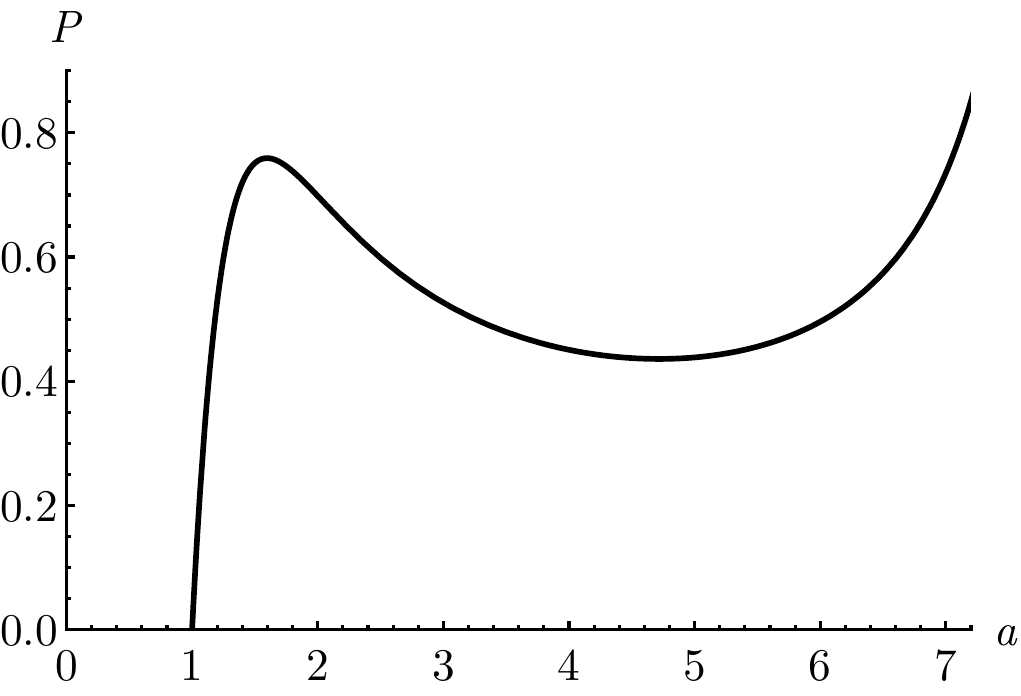}}
	\qquad\qquad
	\subfloat[]{\includegraphics[width=0.43\textwidth]{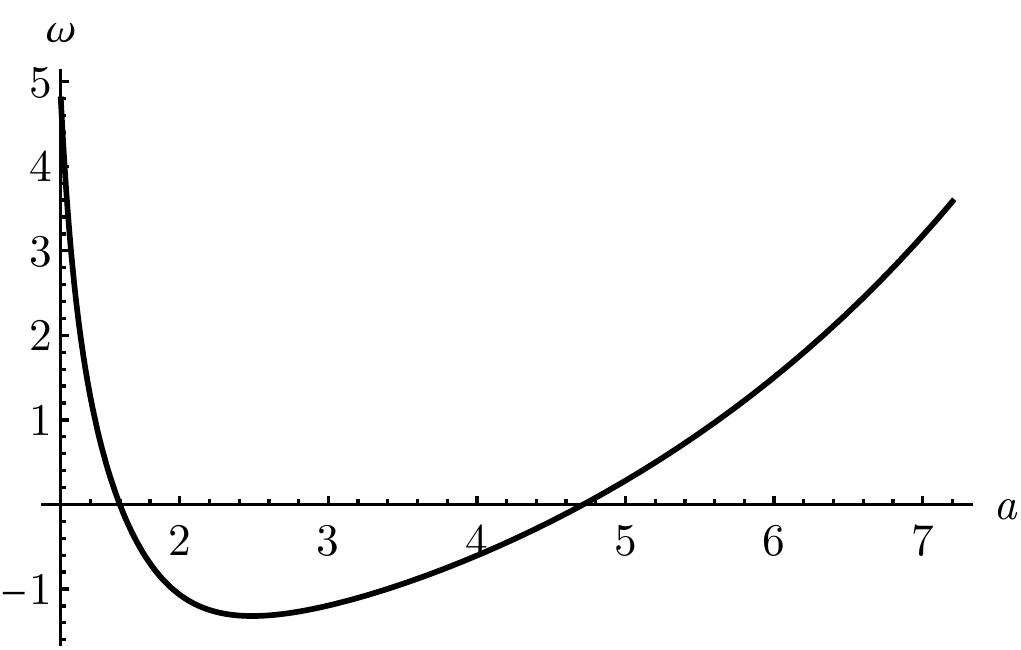}}
	\caption{(a) Pressure–stretch relation for the homogeneous deformation with fixed $N=0$. 
	(b) Variation of $\omega$, the coefficient in the linearized
	equation,  along the loading path \(N=0\).}
	\label{fig:P-a}
\end{figure}

We consider the loading path with fixed axial force \(N=0\). For each
homogeneous radius \(a\), the axial stretch \(\lambda=\lambda(a)\) is
determined implicitly by
\begin{equation}
w_2(a,\lambda(a))-\frac{a w_1(a,\lambda(a))}{2\lambda(a)}=0.
\end{equation}
The first relation in $\eqref{eq:hom}$ then gives the pressure \(P=P(a)\).
Fig.~\ref{fig:P-a}(a) shows the resulting
pressure--stretch relation, which is non-monotonic and exhibits a limit point
associated with the onset of localization.

\subsection{Finite- and infinite-domain weakly nonlinear approximations}

We begin by linearizing about the homogeneous state, a step common to
the finite- and infinite-domain problems. We write
\begin{align}
r(Z)=a+y(Z), \qquad z'(Z)=\lambda(a)+v(Z),
\end{align}
where $y(Z)$ and $v(Z)$ are small. Linearizing \eqref{eq:mem1} and \eqref{eq:mem2} about the homogeneous state and eliminating $v$, we obtain
\begin{align}\label{eq:lin}
y''=\omega(a) y,
\end{align}
where the  coefficient is
\begin{align}\label{eq:omega}
\omega(a)=\frac{\lambda^2 w_{22}(a,\lambda)(aw_{11}(a,\lambda)-w_1(a,\lambda))-a (w_1(a,\lambda)-\lambda w_{12}(a,\lambda))^2}{a \lambda w_2(a,\lambda)w_{22}(a,\lambda)}\Big|_{\lambda=\lambda(a)}.
\end{align}
Here \(w_{ij}=\partial^2w/(\partial\lambda_i\partial\lambda_j)\) for $i,j=1,2$. The variation of \(\omega(a)\) is shown in
Fig.~\ref{fig:P-a}(b).

For the finite-domain problem $Z\in [0,L]$, the boundary conditions \eqref{eq:membc} imply $y'(0)=y'(L)=0$, leading to the modes $y=A\cos(k\pi Z/L)$, $k=1,2,\dots$,
with the dispersion relation $\omega(a)+k^2\pi^2/L^2=0$. The onset of bifurcation is determined by the first mode ($k=1$), yielding
\begin{align}\label{eq:biff}
\omega(a_{\mathrm{cr}})
+\frac{\pi^2}{L^2}=0.
\end{align}

To determine the amplitude, we write
\(a=a_{\mathrm{cr}}+\varepsilon a_1\), where
\(\varepsilon\ll1\) and \(a_1=O(1)\), and apply the finite-domain
weakly nonlinear expansion of Section~\ref{sec:ph}. The solvability
condition at \(O(\varepsilon^{3/2})\) is
\begin{align}\label{eq:amp}
\chi(a_\text{cr}) a_1 A+ A^3=0,
\end{align}
where $\chi(a_\text{cr})$ is determined by the material response at the critical state.   The resulting finite-domain weakly nonlinear approximation is
\begin{align}
r(Z)=a_{\mathrm{cr}}
+\sqrt{-\chi(a_\text{cr}) (a-a_{\mathrm{cr}})}
\cos \frac{\pi Z}{L}+O(|a-a_\text{cr}|).
\end{align}

For the infinite-domain problem $Z\in[0,\infty)$, the linearized equation remains \eqref{eq:lin}. Since no finite-domain end condition selects a discrete mode, the bifurcation condition is
\begin{align}\label{eq:bifcon}
\omega(a^{\infty}_{\mathrm{cr}})=0,
\end{align}
where $a^\infty_{\mathrm{cr}}$ denotes the critical azimuthal stretch for the infinite-domain problem.

Near the critical state, we write $
a = a^\infty_{\mathrm{cr}} + \varepsilon a_1$,
where $\varepsilon\ll1$, and introduce the stretched variable $S=\sqrt{\varepsilon} Z$. Following the infinite-domain scaling in Section~\ref{sec:inf}, we seek solutions of the form
\begin{align}
&r_\infty(Z)=a+\varepsilon y_1(S)+\varepsilon^2 y_2(S)+\cdots,\\
&z'_\infty(Z)=\lambda(a)+\varepsilon v_1(S)+\varepsilon^2 v_2(S)+\cdots,
\end{align}
where \(r_\infty\) and \(z_\infty\) denote the radial and axial
coordinates for the infinite tube.
At $O(\varepsilon^2)$, eliminating \(v_2\)  gives
\begin{align}
y_1''(S)= a_1\omega'(a^{\infty}_{\mathrm{cr}})y_1(S) +\gamma(a^{\infty}_{\mathrm{cr}}) y_1(S)^2,
\end{align}
where \(\gamma(a^\infty_{\mathrm{cr}})\) is determined by the material response at the critical state. The localized solution of this amplitude equation gives the
infinite-domain weakly nonlinear approximation
\begin{align}
r_\infty(Z)=a
-\frac{3(a-a^{\infty}_{\mathrm{cr}})\omega'(a^{\infty}_{\mathrm{cr}})}{2\gamma(a^{\infty}_{\mathrm{cr}})}
\sech^2\Big(\frac{\sqrt{(a-a^{\infty}_{\mathrm{cr}})\omega'(a^{\infty}_{\mathrm{cr}})}}{2}Z\Big)+O(|a-a^\infty_\text{cr}|^2).
\end{align}

As shown in Fig.~\ref{fig:P-a}(a), near the maximum of the pressure–stretch curve, a given pressure can correspond to two homogeneous radii: the larger $a$ corresponds to the center equilibrium used in the finite-domain expansion, and the smaller $a$ to the saddle equilibrium used in the infinite-domain expansion.

\subsection{Length threshold and approximation-error bound}\label{sec:la}

Along the post-bifurcation branch, we quantify the infinite-domain approximation by determining a length threshold $L_*$ and a associated upper bound on the approximation error. We label states along this branch by the far-field radius \(a\).

Let \(A=r(0)-a\) and \(B=r(L)-a\) denote the deviations from the far-field radius at the center and endpoint, respectively.
Following the analysis in Section~\ref{sec:length},
we use the first integral \eqref{eq:fir} to obtain the asymptotic relation
\begin{align}\label{eq:LB}
\sqrt{\omega(a)}L=-\ln B+C(a)+R(B,a),
\end{align}
where $C(a)=\lim_{B\to 0}[\sqrt{\omega(a)}L+\ln B]$ is the load-dependent constant term and $R(B,a)$ is a
remainder satisfying $\lim_{B\to 0}R(B,a)=0$; 
both are obtained by
numerical quadrature of the tube length integral
\(L=\int_{r(0)}^{r(L)} \frac{dr}{r'}\), with $r'$ determined by the first
integral \eqref{eq:fir}. This is the membrane-tube counterpart of
\eqref{eq:l_integral}.

We impose the same one-percent tolerance, measured relative to the
peak deviation \(A\). Let \(b=r_\infty(L)-a\). At the endpoint,
the criterion requires
\begin{align}\label{eq:tube-endpoint-criterion}
|r(L)-r_\infty(L)|=|B-b|\leq0.01A.
\end{align}
For the homoclinic orbit, a similar analysis of the length integral
gives
\begin{align}
\sqrt{\omega(a)}L=-\ln b+C(a)-\ln 2+R_\infty(b,a),
\end{align}
where \(R_\infty(b,a)\to0\) as \(b\to0\). Comparing the two length
relations at the same \(L\) gives \(b=(B/2)e^{R_\infty(b,a)-R(B,a)}\).
At each loading state \(a\), we use this expression and numerical bounds
on the two remainders to obtain an upper bound on \(|B-b|\) in terms of $B$.  To satisfy the endpoint criterion \eqref{eq:tube-endpoint-criterion}
with some room below the tolerance, we choose \(B_0(a)\) so that the
upper bound on \(|B-b|\), increased by ten percent, is at most
\(0.01A\) throughout the range
\begin{align}\label{eq:tube-B-range}
0<B\leq B_0(a).
\end{align}

To express \eqref{eq:tube-B-range} as a length condition, let
\(\bar R(a)\) denote a numerical upper bound on \(R(B,a)\) over
this range. By the length relation \eqref{eq:LB}, \(B\leq B_0(a)\)
is ensured whenever \(L\geq L_*(a)\), where the load-dependent threshold is
\begin{align}\label{eq:tube-length-threshold}
L_*(a)=\frac{-\ln B_0(a)+C(a)+\bar R(a)}{\sqrt{\omega(a)}}.
\end{align}

For \(L\geq L_*(a)\), we estimate the approximation error by adapting
the bootstrap argument in \ref{app:uniform-error}. At each
loading state \(a\), the length relation \eqref{eq:LB} bounds the
endpoint deviation \(B\), and the turning-point relation from the first
integral gives a corresponding bound on the error at \(Z=0\). The
bootstrap argument then controls how this error propagates from the
center toward the endpoint, yielding
\begin{equation}\label{eq:ert}
\sup_{Z\in[0,L]} |r(Z)-r_\infty(Z)|
\leq K(a)e^{-\sqrt{\omega(a)}L}, \qquad L\geq L_*(a),
\end{equation}
where the load-dependent prefactor \(K(a)\) is obtained by numerically
evaluating the bounds used in the bootstrap argument.

\subsection{Numerical computation of fully nonlinear solutions}

We compute the fully nonlinear finite- and infinite-domain solutions
numerically using the exact length integrals obtained from
\eqref{eq:fir}. For the finite tube, equating the first integral at \(Z=0\) and \(Z=L\),
where \(r'=0\), gives an algebraic relation that determines \(r(0)\)
from \(r(L)\). We choose \(r(L)\) so that the length integral from
\(r(L)\) to \(r(0)\) equals the prescribed half-length \(L\). For the
infinite tube, we equate the first integral at \(Z=0\) to its far-field
limit, where \(r_\infty\to a\) and \(r_\infty'\to0\). This gives an
algebraic equation for the center radius \(r_\infty(0)\).

In each case, the radius decreases monotonically away from the center,
so the length integral gives the axial coordinate \(Z\) as a function
of radius.  We evaluate and invert this relation numerically
to obtain \(r(Z)\) and \(r_\infty(Z)\) at prescribed axial positions. We check
convergence separately for each solution by increasing the working
precision and tightening the quadrature and inversion tolerances until
successive values of the radius agree to the required number of digits.

\subsection{Assessment of the infinite-domain approximation and weakly nonlinear validity}

We now use the fully nonlinear solutions to assess whether the length
criterion ensures that the maximum relative error of the infinite-domain
approximation is at most \(1\%\). We also compare each solution with
its weakly nonlinear approximation and examine how the validity range
of the finite-domain expansion varies with the tube length.

For illustration, we take \(L=10\). The bifurcation condition
\eqref{eq:biff} gives the critical stretch and the corresponding pressure
\begin{align}
a_{\mathrm{cr}}=1.6192,
\qquad
P_{\mathrm{cr}}=0.7589.
\end{align}
We use \(P\) as the loading parameter and write
\(\Delta P=P_{\mathrm{cr}}-P\). At each pressure, we find the
far-field radius \(a\) from \eqref{eq:hom} and evaluate the length
threshold \(L_*\) using \eqref{eq:tube-length-threshold}.

\begin{figure}[h!]
	\centering
	\subfloat[]{\includegraphics[width=0.43\textwidth]{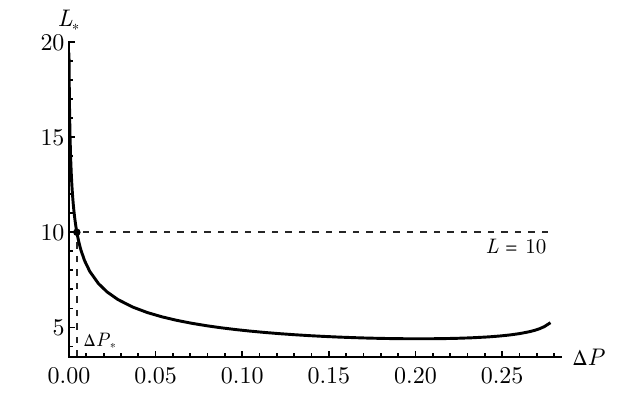}}
	\qquad\qquad
	\subfloat[]{\includegraphics[width=0.43\textwidth]{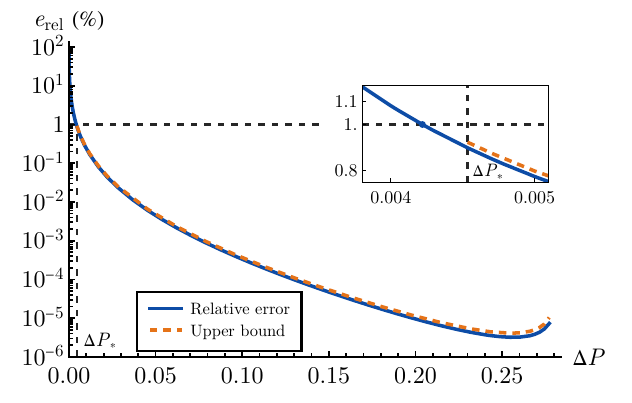}}
	\caption{(a) Length threshold \(L_*\) versus the pressure difference
		\(\Delta P\). The intersection with \(L=10\) predicts a crossover at
		\(\Delta P_*\approx 0.0045\).
		(b) Maximum relative error between the finite- and infinite-domain
		solutions at \(L=10\) and its upper bound from \eqref{eq:ert}. The inset shows
		that the error reaches \(1\%\) at \(\Delta P\approx 0.0042\), close to
		the predicted crossover.}
	\label{fig:Lstar}
\end{figure}

Fig.~\ref{fig:Lstar}(a) shows \(L_*\) as a function of \(\Delta P\). The curve intersects the line \(L=10\) at the predicted crossover
\begin{align}
\Delta P_*\approx 0.0045.
\end{align}
Fig.~\ref{fig:Lstar}(b) shows the maximum relative error between the
finite- and infinite-domain solutions, obtained by dividing the maximum
absolute error by the finite-tube peak deviation \(A=r(0)-a\).
The upper bound follows from \eqref{eq:ert} by the same normalization.  Near bifurcation, the error decreases from about
\(30\%\) at \(\Delta P=0.0001\) to about \(6\%\) at \(\Delta P=0.001\).
It reaches \(1\%\) at \(\Delta P\approx 0.0042\), slightly before the
predicted crossover, and stays below \(1\%\) for
\(\Delta P\geq\Delta P_*\) over the plotted range, in agreement with
the length criterion. The upper bound stays just above the error curve.

\begin{figure}[h!]
	\centering
	\subfloat[]{\includegraphics[width=0.43\textwidth]{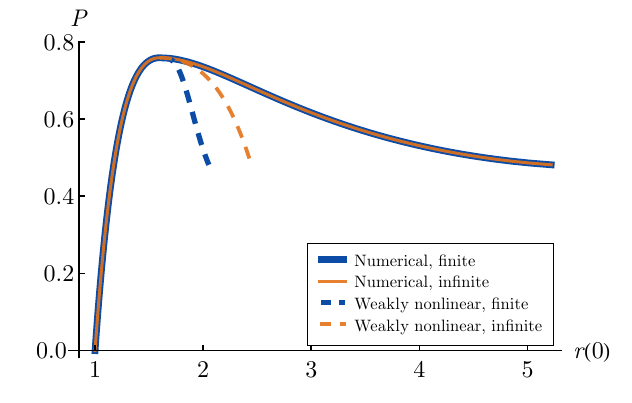}}
	\qquad\qquad
	\subfloat[]{\includegraphics[width=0.43\textwidth]{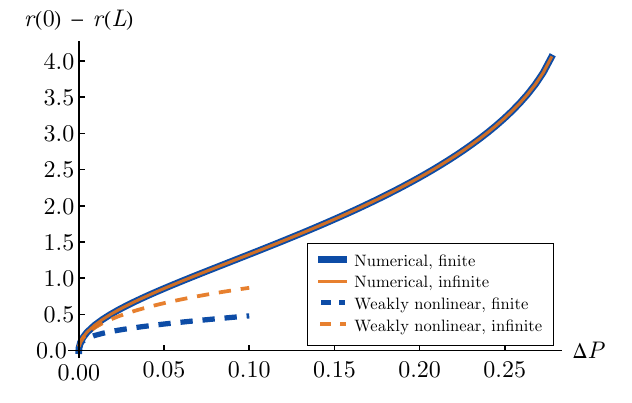}}
	\caption{ 
		(a) Pressure $P$ versus the center radius $r(0)$.
		(b) Bulging amplitude $r(0)-r(L)$ versus the pressure difference
		$\Delta P$.}
	\label{fig:Nz=0}
\end{figure}

Fig.~\ref{fig:Nz=0}(a) shows the pressure--radius relation along the
post-bifurcation branch. Localized bulging emerges near the
pressure maximum, and center radius \(r(0)\) increases as pressure decreases.
The bulging amplitude \(r(0)-r(L)\) also increases with \(\Delta P\),
as shown in Fig.~\ref{fig:Nz=0}(b).

\begin{figure}[h!]
	\centering
	\subfloat[]{\includegraphics[width=0.43\textwidth]{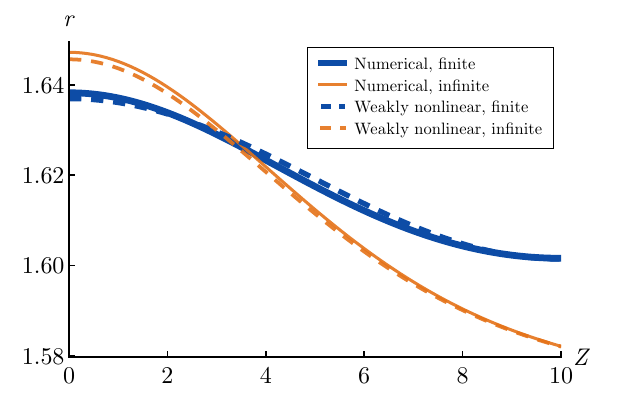}}
	\qquad\qquad
	\subfloat[]{\includegraphics[width=0.43\textwidth]{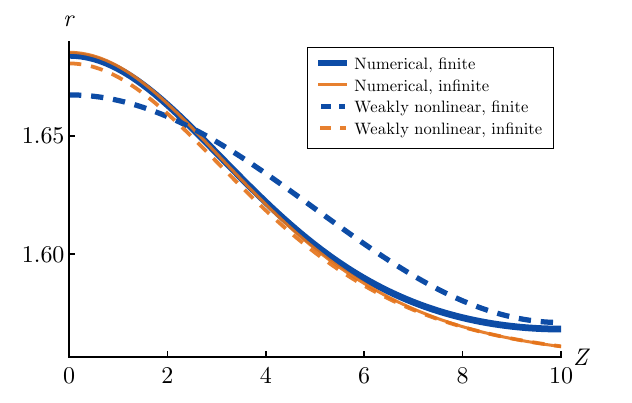}}\\
	\subfloat[]{\includegraphics[width=0.43\textwidth]{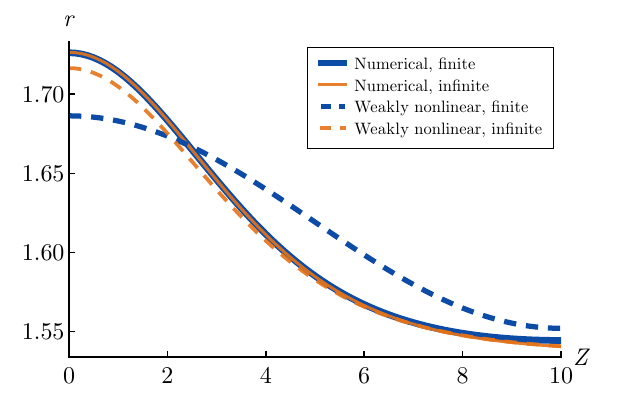}}
	\qquad\qquad
	\subfloat[]{\includegraphics[width=0.43\textwidth]{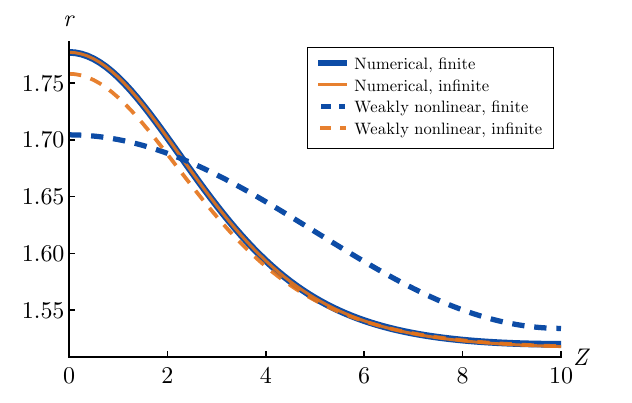}}
	\caption{Comparison of fully nonlinear numerical solutions and weakly nonlinear approximations at four representative states: (a) $\Delta P=0.0001$, (b) $\Delta P=0.001$, (c) $\Delta P=0.0025$, and (d) $\Delta P=0.005$.}
	\label{fig:PP}
\end{figure}

Fig.~\ref{fig:PP} compares the fully nonlinear finite- and infinite-domain
solutions with their corresponding weakly nonlinear approximations at four
representative states. The two fully nonlinear solutions differ visibly near the bifurcation
and approach each other as the loading state moves away from it. At \(\Delta P=0.0025<\Delta P_*\), the two solutions agree closely over
most of the domain, with a visible discrepancy near the tube end.
The maximum relative error falls from \(2.04\%\) at this load to
\(0.78\%\) at \(\Delta P=0.005>\Delta P_*\), below the \(1\%\) tolerance.

\begin{figure}[h!]
	\centering
	\includegraphics[width=0.6\linewidth]{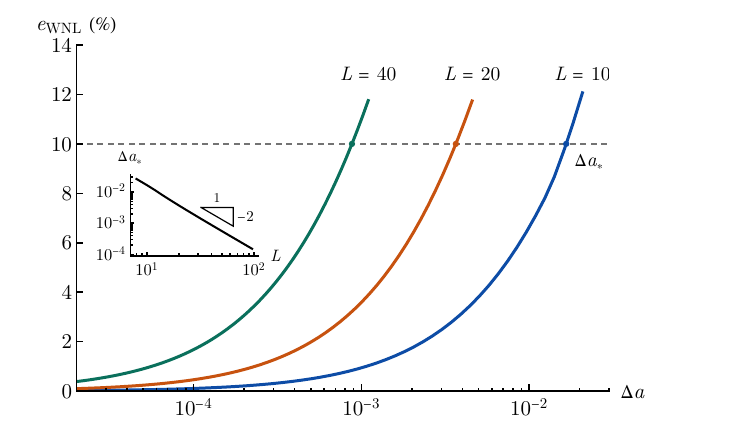}
	\caption{Maximum relative error \(e_{\mathrm{WNL}}\) of the
		finite-domain weakly nonlinear approximation versus
		\(\Delta a\).
		The inset shows the \(L^{-2}\) scaling of \(\Delta a_\star\),
		the increment at which the error first reaches \(10\%\).}
	\label{fig:wnl-validity-scaling}
\end{figure}

As shown in Fig.~\ref{fig:PP}, both weakly nonlinear approximations
accurately reproduce their respective fully nonlinear solutions very
close to bifurcation, confirming the validity of both local expansions
in the immediate neighborhood of bifurcation. As the loading state moves away from the bifurcation point, the
finite-domain weakly nonlinear approximation progressively loses accuracy.
At \(\Delta P=0.005\), the infinite-domain weakly nonlinear approximation
is closer to the finite-domain fully nonlinear solution, although its own
truncation error has also increased.  Fig.~\ref{fig:wnl-validity-scaling}  shows how the validity range of the
finite-domain weakly nonlinear approximation depends on \(L\). Here
\(\Delta a=a-a_{\mathrm{cr}}\) is the increment of the azimuthal stretch from the
finite-domain bifurcation. The value \(\Delta a_\star\)
at which the relative error first reaches \(10\%\) scales approximately
as \(L^{-2}\). The solution comparisons and the observed scaling are
consistent with the validity regimes described in Section~\ref{sec:weakly}.

Overall, the constructive method
gives a load-dependent length threshold for keeping the maximum
relative error of the infinite-domain approximation within \(1\%\),
together with an error bound beyond the crossover. The weakly
nonlinear results show that the validity range of the finite-domain
weakly nonlinear expansion shrinks approximately as \(L^{-2}\).

\section{Application to localized buckling in twisted rods}\label{sec:connection}

To clarify the origin of the prototypical model in Section~\ref{sec:prob},
we consider localized helical buckling in twisted elastic rods
\citep{coyne1990analysis,van2000helical,audoly2010elasticity}.
The prototypical equation follows from the equilibrium of an
\emph{infinite} rod but omits the end correction present in the full finite-rod
equation. We use the constructive method to assess the accuracy of the
infinite-domain approximation for the finite rod.

\subsection{Problem description and governing equations}

\begin{figure}[h!]	
	\centering
	\includegraphics[width=0.8\linewidth]{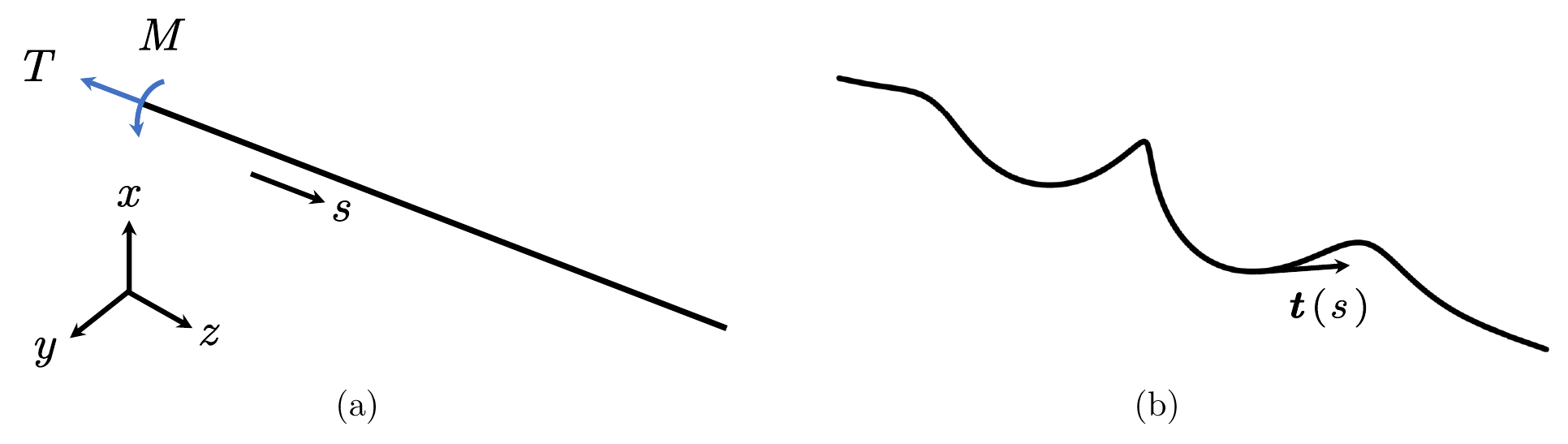}
	\caption{Localized helical buckling of a twisted elastic rod under axial tension
		and twisting moment: (a) reference (straight) configuration, (b) current configuration.}
	\label{fig:twisted-rod}
\end{figure}

Consider a naturally straight elastic rod of total length \(2L\), subjected to an axial tension \(T\) and an applied twisting moment \(M\). The reference centerline is aligned with the \(z\)-axis.  As the applied twisting moment increases, the straight configuration may lose stability and a localized helical buckle emerges, as illustrated in Fig.~\ref{fig:twisted-rod}.

Within the Kirchhoff rod framework, let \(s\in[-L,L]\) denote the arclength
coordinate and \(\bm u(s)\) the centerline position vector. The unit tangent is
\(\bm t(s)=\bm u'(s)\), with the inextensibility constraint $|\bm{u}'(s)|=1$.

Assuming a symmetric cross-section,  the two principal second moments of area coincide. Let \(\theta(s)\) denote the twist angle of the material frame relative to the Bishop frame. The total potential energy then takes the form
\begin{align}
\mathcal E[\bm{u},\theta]
=
\int_{-L}^{L}
\Big[
\frac12 EI|\bm u''(s)|^2
+
\frac12\mu J \theta'(s)^2
-
T\bm{e}_z \cdot \bm{u}'(s)
-
M\theta'(s)\Big]\,ds,
\end{align}
where \(EI\) and \(\mu J\) are the bending and torsional stiffnesses, respectively, and \(\bm{e}_z\) is the unit vector along the undeformed rod axis.

Because the Bishop frame varies with the centerline, the applied twisting
moment contributes the additional virtual-work term
\begin{align}\label{eq:vw}
\delta W
=-\int_{-L}^{L}
M[\bm{u}'(s)\times  \bm{u}''(s)]\cdot\delta \bm {u}'(s)\,ds .
\end{align}
The equilibrium configurations are therefore determined by the constrained variational principle $\delta\mathcal E-\delta W=0$, subject to $|\bm{u}'(s)|=1$.

To derive the governing equations, we parameterize the unit tangent vector using the spherical angles \(\phi(s)\) and \(\psi(s)\):
\begin{align}\label{eq:up}
\bm{u}'(s)=\cos\psi(s)\sin\phi(s)\bm{e}_x+\sin\psi(s)\sin\phi(s)\bm{e}_y+\cos\phi(s)\bm{e}_z,
\end{align}
where \(\bm e_x\), \(\bm e_y\), and \(\bm e_z\) are the Cartesian basis vectors.
Substituting this representation into the constrained variational principle  and taking independent variations with respect to \(\theta\), \(\phi\), and \(\psi\) yields
\begin{align}
&\mu J\theta'=M,\label{eq:eqtheta}\\
&EI(\phi''-\cos\phi\sin\phi\psi'^2)-T \sin\phi +M\sin\phi \psi' =0,\label{eq:eq1}\\
&\sin\phi[EI(\sin\phi\psi''+2\cos\phi \psi'\phi') -M\phi']=0,\label{eq:eq2}
\end{align}
where the primes denote differentiation with respect to $s$. The associated natural boundary conditions are
\begin{align}\label{eq:bcc}
\phi'=0,
\qquad
\sin^2\phi \psi'=0
\qquad \text{at } s=\pm L.
\end{align}

\subsection{Scalar rod equation  and its connection to the prototypical model}

The twisting-angle equation \eqref{eq:eqtheta} is decoupled and does not
enter the centerline problem. Integrating \eqref{eq:eq2} once and using the end condition $\eqref{eq:bcc}$ gives
\begin{align}\label{eq:psi}
EI\sin^2\phi\,\psi'=M\big(\cos\phi(L)-\cos\phi(s)\big).
\end{align}
Multiplying \eqref{eq:eq1} by \(\phi'\) and \eqref{eq:eq2} by \(\psi'\),
integrating their sum, and using the end conditions \eqref{eq:bcc} to evaluate
the integration constant, we obtain the second first integral
\begin{align}\label{eq:second_integral}
\frac{EI}{2}(\phi'^2+\sin^2\phi \psi'^2 )
=T\big(\cos\phi(L)-\cos\phi(s)\big).
\end{align}

Setting \(y=1-\cos\phi\) and \(B=y(L)\), and eliminating \(\psi'\) in \eqref{eq:second_integral}
using \eqref{eq:psi}, we obtain
\begin{align}\label{eq:rod-finite-integral}
\frac12 y'^2=q(y-B)[y(2-y)-2\lambda(y-B)],
\end{align}
where tension parameter $q$ and twist parameter $\lambda$ are defined by 
\begin{align}\label{eq:qlambda}
q=\frac{T}{EI}, \qquad \lambda=\frac{M^2}{4EIT}.
\end{align}
Differentiating \eqref{eq:rod-finite-integral} gives the scalar equilibrium
equation for \(y\) in a finite rod,
\begin{align}\label{eq:rod-finite}
y''=4q(1-\lambda)y-3qy^2+2qB(y+2\lambda-1).
\end{align}
For a localized buckle symmetric about \(s=0\), symmetry and the end
condition \(\phi'(L)=0\) give the boundary conditions
\begin{align}
y'(0)=0, \qquad y'(L)=0.
\end{align}

For an infinite rod, the centerline approaches the straight
configuration as \(s\to\infty\), so \(y,y'\to0\). The far-field value of \(y\) therefore gives \(B=0\). The scalar equilibrium
equation \eqref{eq:rod-finite} then reduces to
\begin{align}\label{eq:rod-prototypical}
y''=4q(1-\lambda)y-3qy^2.
\end{align}
This is the prototypical localization equation \eqref{eq:orig}, with
\(\alpha(\lambda)=4q(1-\lambda)\) and \(\beta(\lambda)=-3q\).
From \eqref{eq:yinf}, its localized solution centered at \(s=0\) is
\begin{align}\label{eq:infinite-rod}
y_\infty(s)=2(1-\lambda)
\sech^2 (\sqrt{q(1-\lambda)} s).
\end{align}
The localized branch thus bifurcates from the straight configuration
at \(\lambda=1\), and the amplitude of the localized solution increases
as \(\lambda\) decreases.

On a finite domain, the prototype model retains the finite-rod
boundary conditions \(y'(0)=y'(L)=0\) but omits the end correction
\(2qB(y+2\lambda-1)\). The resulting equation
\eqref{eq:rod-prototypical} therefore has the same form on both domains.
This simplification provides a simple setting for assessing the infinite-domain
approximation.

\subsection{Assessment of the infinite-domain approximation}

We now apply the constructive method  to
the full finite-rod equation \eqref{eq:rod-finite}. As in Section~\ref{sec:length}, we use the length integral to estimate
the end value \(B=y(L)\). The bootstrap estimate also accounts for
the end correction \(2qB(y+2\lambda-1)\) in the error equation.
With this correction included, the rod problem can still be analyzed
analytically, as shown in \ref{app:rod-estimates}.

For \(q>0\) and \(0<\lambda<1\), let \(m(\lambda)=\max(\lambda,1-\lambda)\).
The resulting length threshold is
\begin{align}\label{eq:rod-length-threshold}
L_*(q,\lambda)
=\frac{\ln(1870\,m(\lambda))}
{2\sqrt{q(1-\lambda)}}.
\end{align}
For \(L\geq L_*(q,\lambda)\), the maximum error satisfies
\begin{align}\label{eq:rod-error-bound}
\sup_{0\leq s\leq L}|y(s)-y_\infty(s)|
\leq 34 (1-\lambda)m(\lambda)
e^{-2L\sqrt{q(1-\lambda)}}.
\end{align}
Since \(y(0)\geq 2(1-\lambda)\), the maximum error relative to the
finite-rod peak \(y(0)\) is at most \(17/1870\approx 0.909\%<1\%\)
for \(L\geq L_*\).

\begin{figure}[h!]
	\centering
	\subfloat[]{\includegraphics[width=0.43\textwidth]{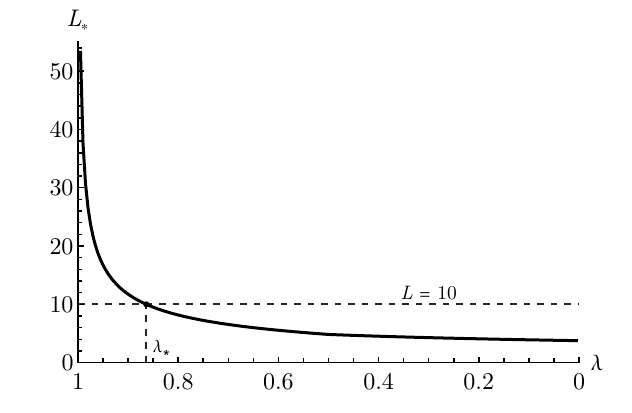}}
	\qquad\quad
	\subfloat[]{\includegraphics[width=0.45\textwidth]{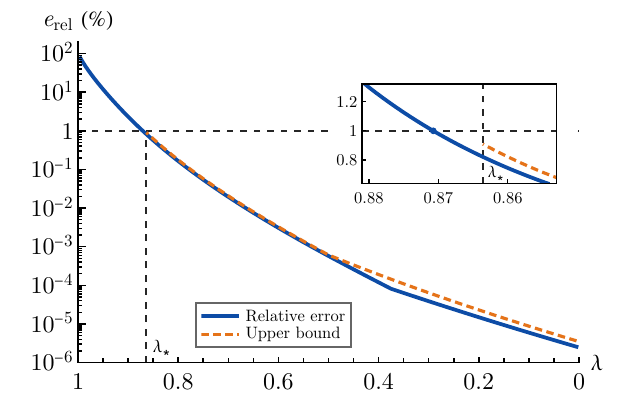}}
	\caption{(a) Length threshold \(L_*\) versus \(\lambda\) for \(q=1\),
		with a predicted crossover at \(\lambda_*\approx0.864\) for \(L=10\).
		(b) Maximum relative error between the finite- and infinite-rod solutions
		and its upper bound at \(L=10\). The inset shows that the maximum relative error reaches \(1\%\)
		at \(\lambda\approx0.871\), close to the predicted crossover.
	}
	\label{fig:rod-criterion}
\end{figure}

For illustration, we take \(q=1\) and \(L=10\) and vary \(\lambda\)
along the localized post-buckling branch. At each load, we compute the
finite-rod solution from the first integral \eqref{eq:rod-finite-integral}
and compare it with the analytical infinite-rod solution \eqref{eq:infinite-rod}.
Fig.~\ref{fig:rod-criterion} (a) shows \(L_*\) as a function of
\(\lambda\). Its intersection with \(L=10\) gives the predicted crossover
\begin{align}
\lambda_*\approx0.864.
\end{align}
Fig.~\ref{fig:rod-criterion}(b) shows the maximum relative error and
its upper bound from \eqref{eq:rod-error-bound}. As \(\lambda\)
decreases, the error reaches \(1\%\) at \(\lambda\approx0.871\),
slightly before the predicted crossover. It stays below \(1\%\) for
\(\lambda\leq\lambda_*\) over the plotted range, in agreement with the
length criterion. The upper bound lies above the error curve and
follows it closely.

\begin{figure}[h!]
	\centering
	\subfloat[]{\includegraphics[width=0.43\textwidth]{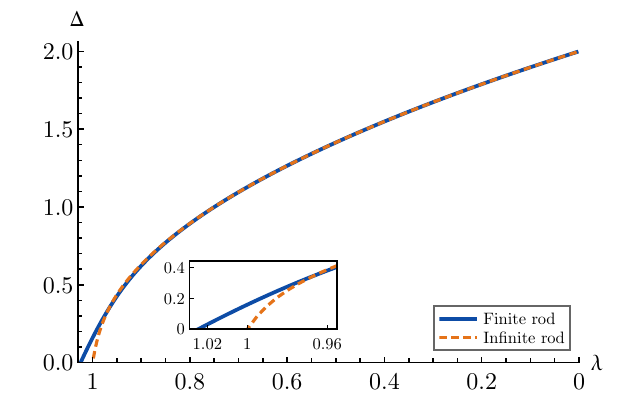}}
	\qquad\quad
	\subfloat[]{\includegraphics[width=0.43\textwidth]{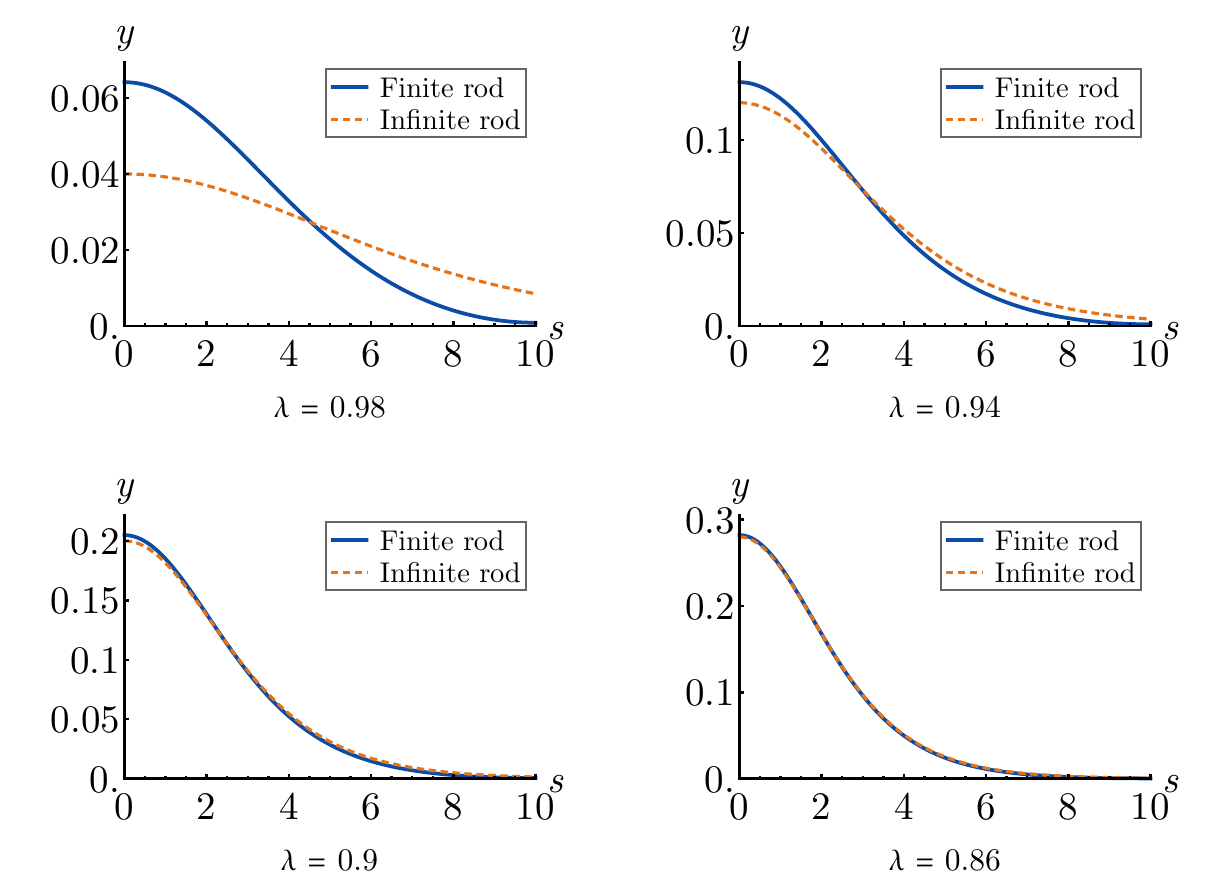}}
	\caption{Finite- and infinite-rod solutions for \(q=1\) and \(L=10\).
		(a) End-shortening \(\Delta\) versus the twist parameter \(\lambda\).
		The inset shows the two branches near bifurcation.
		(b) Localized solutions \(y(s)\) at \(\lambda=0.98,\ 0.94,\ 0.90,\ 0.86\).
	}
	\label{fig:RR}
\end{figure}

Fig.~\ref{fig:RR}(a) shows the end-shortening
\(\Delta=L-u_3(L)=\int_0^L y(s)\,ds\) along the post-buckling branch.
The two loading curves differ visibly near bifurcation, as shown in
the inset. The finite rod bifurcates at \(\lambda\approx1.025\) and
the infinite rod at \(\lambda=1\). As \(\lambda\) decreases, the
end-shortening increases and the two curves nearly coincide.

Fig.~\ref{fig:RR}(b) compares the finite- and infinite-rod solutions
at four representative loading states. At \(\lambda=0.98\), the finite-rod
solution has a higher peak and a smaller end value, with a maximum
relative error of about \(38\%\). As \(\lambda\) decreases, the error
falls to \(8.4\%\) at \(\lambda=0.94\) and \(2.3\%\) at \(\lambda=0.90\).
At \(\lambda=0.86\), just below the predicted crossover
\(\lambda_*\approx0.864\), the solutions are nearly indistinguishable
throughout the domain, with a maximum relative error of \(0.75\%\).

Overall, the constructive method identifies when the infinite-rod solution
accurately approximates the finite-rod solution and gives an upper bound
on the error along the entire rod.

\section{Conclusion}\label{sec:con}

This work has developed a constructive method for quantitatively assessing
the infinite-domain approximation for elastic localization in finite
structures. For the prototype model, the first integral
\eqref{eq:first-integral} gives the length relation
\eqref{eq:length-amplitude-explicit}, which determines the endpoint
deviation from the far-field state. The turning-point relation
\eqref{eq:A-of-B} converts this deviation into the error at the
localization center. A bootstrap argument controls the propagation of
this error across the domain, yielding the length criterion
\(L\geq L_*(\lambda)\), with the load-dependent threshold given by
\eqref{eq:physical-length-threshold}, and the bound
\eqref{eq:uniform-physical-error} on the maximum approximation error.

The length criterion and error bound allow the infinite-domain approximation
to be assessed before the finite-domain solution is computed, whereas direct
numerical comparison gives the error for a particular length and loading
state. At each load, the constructive method reduces the assessment to two
quantities, the length threshold and the error prefactor. For a given
length, the threshold predicts whether the infinite-domain approximation achieves the
prescribed accuracy. For all lengths at or above the threshold, the
prefactor and the decay rate from the far-field linearization give an
explicit exponential upper bound on the maximum error throughout the domain.

The length threshold increases rapidly near bifurcation. For a structure
of fixed length, the infinite-domain approximation therefore loses
accuracy as bifurcation is approached. The crossover occurs at
\(\varepsilon=O(L^{-2})\), where \(\varepsilon\) measures the distance
in load from bifurcation. The same crossover also accounts for the change
between the finite- and infinite-domain weakly nonlinear descriptions
and for the narrowing validity range of the finite-domain expansion as
the structural length increases.

For the membrane tube, numerical evaluation of the model-dependent
estimates yields a load-dependent length threshold and an upper bound
on the approximation error. Comparisons with the fully nonlinear
solutions also support the predicted validity regimes of the two weakly
nonlinear approximations. For the twisted rod, the threshold and error
bound are obtained analytically with the end correction included in the
full finite-rod equation. In both applications, the predicted crossover
lies close to that obtained numerically, and the upper bound closely
tracks the maximum error beyond the crossover.

The present method is limited to one-dimensional localization problems
with translational invariance. Future work may consider problems without
this invariance, such as localized buckling in spherical shells
\citep{audoly2020localization}, together with higher-dimensional
localization and the effects of geometric or material imperfections.

All numerical computations reported in this paper were performed using
\textit{Mathematica} \citep{wolfram2026mathematica}, with the 
code available on GitHub (\url{https://github.com/xiangyudgut/elastic-localization-length-criterion}).

\section*{Acknowledgements}
 
This work was supported by the National Natural Science Foundation of China (Grant No 12402068), Guangdong
 Basic and Applied Basic Research Foundation (Grant No 2023A1515111141) and Youth S\&T Talent Support Programme of Guangdong Provincial
 Association for Science and Technology (No SKXRC2025437).
 	
 \appendix
 
\section{Quantitative estimate of the length integral}
\label{app:length-evaluation}

In this appendix, we first show that the remainder \(R(B)\) in the
length--amplitude relation \eqref{eq:length-amplitude-explicit} tends
to zero as the endpoint value \(B\) tends to zero. We then prove that
\(R(B)\) is strictly increasing and use these two properties to derive
an explicit bound on the remainder.

Using the splitting \(\ell=I_1+I_2\) in \eqref{eq:I1I2-main} and
\(B=\delta^2\), we write
\begin{align}\label{eq:RofB}
R(B)
&=\Big(I_1-\ln\frac{2}{\delta}\Big)
+\Big(I_2-\ln\frac{6}{\delta}\Big).
\end{align}
It suffices to show that each term tends to zero as \(B\to0^+\).

For \(I_1\), we use \(V(u)=-u^2/2+u^3/3\) to factor the radicand
at the lower turning point:
\begin{align}
2\bigl(V(B)-V(u)\bigr)
&=(u^2-B^2)
\Big[1-\frac23\Big(u+\frac{B^2}{u+B}\Big)\Big].
\end{align}
For \(B<u\leq\delta\), the term subtracted from \(1\) lies between
\(0\) and \(u\). We use the elementary inequality
\begin{align}\label{eq:app-inverse-sqrt-bound}
0\leq \frac{1}{\sqrt{1-x}}-1\leq 2x,
\qquad 0\leq x\leq \frac12.
\end{align}
For \(\delta\leq \frac{1}{2}\), it follows that
\begin{align}\label{eq:I1B}
0\leq I_1-\int_B^\delta\frac{du}{\sqrt{u^2-B^2}}
&\leq 2\int_B^\delta\frac{u\,du}{\sqrt{u^2-B^2}}
=2\sqrt{\delta^2-B^2}\leq 2\delta.
\end{align}
Since \(B=\delta^2\), direct integration gives
\begin{align}
\int_B^\delta\frac{du}{\sqrt{u^2-B^2}}
&=\ln\frac{1+\sqrt{1-\delta^2}}{\delta}
=\ln\frac{2}{\delta}
+\ln\frac{1+\sqrt{1-\delta^2}}{2}.
\end{align}
The last term tends to zero as \(B\to0^+\). Together with the error
estimate \eqref{eq:I1B}, this gives \(I_1-\ln(2/\delta)\to0\) as
\(B\to0^+\).

For \(I_2\), we use \(V(B)=V(A)\) to factor the same radicand
at the upper turning point:
\begin{align}
2\bigl(V(B)-V(u)\bigr)
&=u^2(1-u/A)
\Big[1+\frac{2V(B)}{u^2}
\Big(1+\frac{u}{A}+\frac{u^2}{A^2}\Big)\Big].
\end{align}
Since \(0<-2V(B)\leq B^2\), the bracketed factor lies between
\(1-3B\) and \(1\) on \(\delta\leq u\leq A\), where we used
\(\delta^2=B\). For \(3B\leq1/2\), the inequality \eqref{eq:app-inverse-sqrt-bound} gives
\begin{align}\label{eq:I2B}
0\leq I_2-\int_\delta^A\frac{du}{u\sqrt{1-\frac{u}{A}}}
&\leq 6B\int_\delta^A\frac{du}{u\sqrt{1-\frac{u}{A}}}.
\end{align}
Direct integration gives
\begin{align}
\int_\delta^A\frac{du}{u\sqrt{1-\frac{u}{A}}}
&=\ln\frac{6}{\delta}
+\ln \Big[\frac{A}{6}
\Big(1+\sqrt{1-\delta/A}\Big)^2\Big].
\end{align}
The second term tends to zero as \(B\to0^+\), since
\(A\to3/2\) and \(\delta=\sqrt{B}\to0\).  Since \(B\ln(6/\delta)\to0\) as well, the estimate \eqref{eq:I2B} yields \(I_2-\ln(6/\delta)\to0\) as
\(B\to0^+\).

Since both terms on the right-hand side of \eqref{eq:RofB} tend to zero,
we conclude that
\begin{align}
\lim_{B\to0^+} R(B)=0.
\end{align}
Thus \(\ln 12\) is the constant term in the length--amplitude relation
 \eqref{eq:length-amplitude-explicit} as \(B\to0^+\).

To bound \(R(B)\) away from the limit \(B\to0^+\), we next show that
it is strictly increasing on \(0<B<1\). We use \(H=V(B)=V(A)\) as
the parameter and define
\begin{align}\label{eq:Jk}
J_k(H)=\int_B^A u^k\sqrt{2(H-V(u))}\,du.
\end{align}
Differentiating with respect to \(H\) gives
\begin{align}\label{eq:Jkp}
J_k'(H)=\int_B^A\frac{u^k}{\sqrt{2(H-V(u))}}\,du.
\end{align}
The endpoint terms vanish because \(V(A)=V(B)=H\).
In particular, \(J_0'(H)=\ell\).

Since \(V(A)=V(B)=H\), we have
\begin{align}
\int_B^A \frac{d}{du}\big[u^k\sqrt{2(H-V(u))}\big]\,du
&=u^k\sqrt{2(H-V(u))}\big|_B^A=0.
\end{align}
Expanding the derivative and taking \(k=0,1,2\) gives, respectively,
\begin{align}\label{eq:Jp}
J_2' &= J_1', \qquad
J_3' = J_2'+J_0, \qquad
J_4' = J_3'+2J_1.
\end{align}
For \(k=0,1\), using \(2(H-V(u))=2H+u^2-\tfrac23u^3\)
in \eqref{eq:Jk} gives
\begin{align}\label{eq:Jpadd}
J_0 = 2H\ell + J_2' - \tfrac23 J_3', \qquad J_1 = 2H J_1' + J_3' - \tfrac23 J_4'.
\end{align}
Substituting \eqref{eq:Jp} into \eqref{eq:Jpadd} and collecting terms,
we obtain
\begin{align}\label{eq:J-identities}
5J_0 &= J_1'+6H\ell, \qquad
7J_1 = (1+6H)J_1'+J_0.
\end{align}

We differentiate both identities in \eqref{eq:J-identities} with
respect to \(H\) and use \(J_0'=\ell\). Eliminating \(J_1''\) between
the resulting equations and substituting \(J_1'=5J_0-6H\ell\)
from the first identity in \eqref{eq:J-identities} gives
\begin{align}\label{eq:app-length-derivative}
\frac{d\ell}{dH}
&=\frac{5J_0(H)}{36(-H)(H+\tfrac16)}.
\end{align}

Using \(H=-B^2(3-2B)/6\), we differentiate
\(R(B)=\ell(B)-\ln(12/B)\) to obtain
\begin{align}\label{eq:dRdB}
\frac{dR}{dB}
&=\frac1B+\frac{6(1-B)}{B(3-2B)}
H\frac{d\ell}{dH}.
\end{align}
Since \(6(1-B)/[B(3-2B)]>0\) for \(0<B<1\), a lower bound on
\(H\,d\ell/dH\) gives a lower bound on \(dR/dB\).
Using \eqref{eq:app-length-derivative}, \(J_0'=\ell\), and
\(J_0(-1/6)=0\), we obtain
\begin{align}
H\frac{d\ell}{dH}
&=-\frac{5}{36}\frac{J_0(H)}{H+1/6}
=-\frac{5}{36}\frac{1}{H+1/6}\int_{-1/6}^{H}\ell(t)\,dt.
\end{align}
Since \(-1/6<H<0\) and \(J_0(H)>0\), \eqref{eq:app-length-derivative}
shows that \(\ell\) increases with \(H\). The average of \(\ell\) over
\([-1/6,H]\) therefore increases with \(H\) and is bounded above by
its limit as \(H\to0^-\). Since
\(J_0(0)=\int_0^{3/2}u\sqrt{1-\tfrac23u}\,du=3/5\), this gives
\begin{align}
H\frac{d\ell}{dH}
&\geq -\frac{5}{36}\frac{J_0(0)}{1/6}=-\frac12.
\end{align}
Substituting this bound into \eqref{eq:dRdB} yields
\begin{align}\label{eq:app-remainder-monotonicity}
\frac{dR}{dB}
&\geq \frac{1}{B}-\frac{3(1-B)}{B(3-2B)}
=\frac{1}{3-2B}>0,
\qquad 0<B<1.
\end{align}
Thus \(R(B)\) is strictly increasing on \(0<B<1\).

For the range of \(B\) needed in Section~\ref{sec:criterion},
evaluating \(R(B)\) at \(B=0.03\) and using monotonicity gives
\begin{align}\label{eq:app-total-bound}
0<R(B)\leq R(0.03)\leq 0.012,
\qquad 0<B\leq0.03.
\end{align}

\section{Estimate of the approximation error}
\label{app:uniform-error}

This appendix establishes bounds on the maximum difference between the finite- and infinite-domain solutions of the prototype model and determines its leading-order asymptotic behavior.

 Define the error function by
\begin{equation}
w(s)=u(s)-u_\infty(s),
\end{equation}
where $u_\infty(s)=\frac32\sech^2\frac{s}{2}$. Subtracting the two equations satisfied by $u$ and $u_\infty$ gives
\begin{equation}\label{eq:refined-error-equation}
w''
=(1-u-u_\infty)w=
(1-2u_\infty)w-w^2=\Big(1-3\sech^2\frac{s}{2}\Big)w-w^2.
\end{equation}

Consider first the homogeneous linearized equation
\begin{equation}\label{eq:variational-equation}
v''=
\left(1-3\sech^2\frac{s}{2}\right)v.
\end{equation}
A convenient fundamental pair of solutions is given by
\begin{align}
&v_1(s)
=
\frac{24-\cosh s (7+\cosh s)-15s\tanh(s/2)
}{8(1+\cosh s)
},\label{eq:phi-solution}
\\
&v_2(s)
=
2\sech^2\frac{s}{2}\tanh\frac{s}{2},
\label{eq:theta-solution}
\end{align}
which satisfy
\begin{equation}
v_1(0)=1,\qquad v_1'(0)=0,
\qquad
v_2(0)=0,\qquad v_2'(0)=1.
\end{equation}
Their Wronskian is therefore equal to one. For \(s>0\), the solutions satisfy
\begin{equation}\label{eq:fundamental-solution-bounds}
|v_1(s)|
\leq 1+\frac{e^s}{16},\quad v_2(s)>0.
\end{equation}

The Green function associated with \eqref{eq:variational-equation} is
\begin{equation}\label{eq:green-function}
G(s,t)
=
v_2(s)v_1(t)-v_1(s)v_2(t),
\qquad
0\leq t\leq s.
\end{equation}
For \(0<t<s\), the unit-Wronskian relation gives
\begin{equation}
G(s,t)
=
v_2(s)v_2(t)
\int_t^s\frac{d\xi}{v_2^2(\xi)}
\geq 0,
\end{equation}
with the endpoint cases following by continuity.
For fixed $t$, $G(t,t)=0$ and $\frac{\partial G}{\partial s}(t,t)=1$. Applying variation of constants to
\begin{equation}
\frac{\partial^2G}{\partial s^2}-G
=
-3\sech^2\frac{s}{2}\,G
\end{equation}
gives
\begin{equation}
G(s,t)
=
\sinh(s-t)
-
3\int_t^s
\sinh(s-\xi)\sech^2\frac{\xi}{2} G(\xi,t)\,d\xi.
\end{equation}
The integral is nonnegative, and hence
\begin{equation}\label{eq:green-bound}
0\leq G(s,t)
\leq
\sinh(s-t)
\leq
\frac12e^{s-t},
\qquad
0\leq t\leq s.
\end{equation}

Applying variation of constants to \eqref{eq:refined-error-equation} and using $w'(0)=0$ give
\begin{equation}\label{eq:w-variation}
w(s)
=
w(0)v_1(s)
-
\int_0^sG(s,t)w^2(t)\,dt.
\end{equation}
The relation between \(A\) and \(B\), together with \eqref{eq:BB}, gives
\begin{equation}\label{eq:w0-bound}
|w(0)|=\frac32-A
\leq \frac23 B^2
\leq \frac23(12.2)^2 e^{-2\ell}.
\end{equation}

Let \(h(s)=1+e^s/16\) and suppose that
\begin{align}
|w(t)|\leq 1.02|w(0)|h(t),
\qquad 0\leq t\leq s.
\end{align}
Using \eqref{eq:fundamental-solution-bounds},
\eqref{eq:green-bound}, and \eqref{eq:w-variation}, we obtain
\begin{align}
|w(s)|
&\leq |w(0)|h(s)
+\frac{1.02^2}{2}w(0)^2e^s
\Big(1-e^{-s}+\frac{s}{8}+\frac{e^s-1}{256}\Big).
\label{eq:w-refined-intermediate}
\end{align}
Since \(e^s/h(s)\leq16\), the ratio of the second term to the first
in \eqref{eq:w-refined-intermediate} is bounded by
\(\tfrac{1.02^2}{2}Q(\ell)\), where
\begin{equation}
Q(\ell)
=\frac{32}{3}(12.2)^2e^{-2\ell}
\Big(1+\frac{\ell}{8}+\frac{e^\ell}{256}\Big).
\end{equation}
The function \(Q\) is decreasing, with \(Q(\ell_*)<0.028\) at the
threshold \(\ell_*\approx6.12\). Thus,
\begin{align}
|w(s)|
&\leq \Big(1+\frac{1.02^2}{2}Q(\ell)\Big)|w(0)|h(s)
\leq 1.015 |w(0)|h(s),
\end{align}
which closes the bootstrap argument. It follows from
\eqref{eq:w0-bound} that
\begin{align}
\sup_{0\leq s\leq\ell}|w(s)|
&\leq 1.015\,\frac23(12.2)^2
\Big(\frac1{16}+e^{-\ell_*}\Big)e^{-\ell}
\leq 6.6e^{-\ell}.
\end{align}

A lower bound follows directly from the
endpoint values. By \eqref{eq:length-amplitude-explicit} and
\eqref{eq:length-remainder-bound}, \(B\geq 12e^{-\ell}\), while
\(u_\infty(\ell)\leq 6e^{-\ell}\). Hence
\begin{align}
|u(\ell)-u_\infty(\ell)|=B-u_\infty(\ell)\geq 6e^{-\ell}.
\end{align}
Thus, the maximum difference satisfies
\begin{align}
6e^{-\ell}
\leq \sup_{s\in[0,\ell]}|u(s)-u_\infty(s)|
\leq 6.6e^{-\ell},
\qquad \ell\geq\ell_*.
\end{align}

A refinement of the preceding argument gives the leading asymptotic
prefactor. The length--endpoint relation \eqref{eq:length-amplitude-explicit} and the turning-point relation \eqref{eq:A-of-B}
yield
\begin{equation}
B=12e^{-\ell}(1+o(1)),
\qquad
w(0)=A-\frac32=-96e^{-2\ell}(1+o(1)).
\end{equation}
Moreover,
\begin{equation}
\sup_{s\in[0,\ell]}|v_1(s)|
=
\frac1{16}e^\ell(1+o(1)).
\end{equation}
Since $|w(0)|=O(e^{-2\ell})$, the preceding estimates give
\begin{align}
\sup_{s\in[0,\ell]}
\Big|\int_0^s G(s,t)w^2(t)\,dt\Big|
&\leq |w(0)|^2e^\ell
\Big(1+\frac{\ell}{8}+\frac{e^\ell}{256}\Big)
=O(e^{-2\ell})
=o(e^{-\ell}).
\end{align}
Together with the leading-order estimate of $w(0)v_1(s)$, this yields
\begin{equation}
\sup_{s\in[0,\ell]}|w(s)|
\leq
6e^{-\ell}(1+o(1)).
\end{equation}
At the endpoint,
\begin{equation}
|w(\ell)|
=
B-u_\infty(\ell)
=
12e^{-\ell}(1+o(1))-6e^{-\ell}(1+o(1))
=
6e^{-\ell}(1+o(1)),
\end{equation}
which provides the matching lower bound. Therefore,
\begin{equation}\label{eq:uniform-error-asymptotic}
\sup_{s\in[0,\ell]}|u(s)-u_\infty(s)|
=
6e^{-\ell}(1+o(1)),
\qquad \ell\to\infty.
\end{equation}

Although explicit formulas are available for the prototype model, the estimate is built from the error representation \eqref{eq:w-variation}: a bound on the initial error $w(0)$, bounds on the solution $v_1$ of the linearized equation and on the Green function $G$, and bootstrap control of the nonlinear term. The bootstrap argument is therefore not tied to explicit solution formulas
and can be adapted to fully nonlinear problems by evaluating the required
model-dependent bounds numerically.

\section{Length threshold and error bound for twisted rods}\label{app:rod-estimates}

This appendix derives the length threshold \eqref{eq:rod-length-threshold}
and the bound \eqref{eq:rod-error-bound} on the maximum error for the
twisted rod.

Let \(A=y(0)\) and \(B=y(L)\). Evaluating
\eqref{eq:rod-finite-integral} at the center gives
\begin{align}
A=1-\lambda+\sqrt{(1-\lambda)^2+2\lambda B}.
\end{align}
Integrating the same first integral from the center to the end gives
\begin{equation}\label{eq:rod-app-length}
L=\frac{1}{\sqrt{2q}}\int_B^A
\frac{dy}{\sqrt{(y-B)(A-y)(y+A-2(1-\lambda))}}.
\end{equation}
Extracting the logarithmic term  gives
\begin{align}
2L\sqrt{q(1-\lambda)}
=-\ln B+\ln \big(32(1-\lambda)^2\big) +R(B),
\end{align}
where \(R(B)\to0\) as $B\to 0^+$ for fixed \(0<\lambda<1\). 
For the end values used below, the remainder satisfies
\begin{align}\label{eq:rod-remainder-bound}
R(B)\leq \ln\frac{17}{16},\qquad
0<B\leq\frac{(1-\lambda)^2}{50\,m(\lambda)},
\end{align}
where \(m(\lambda)=\max(\lambda,1-\lambda)\).

To impose the endpoint criterion \(|B-y_\infty(L)|\leq 0.01A\), we
include the same ten-percent margin and require
\(1.1\,m(\lambda)B \leq 0.02(1-\lambda)^2\). This gives
\begin{align}
B \leq B_0, \qquad
B_0 = \frac{(1-\lambda)^2}{55\,m(\lambda)}.
\end{align}
Since \(B\leq B_0\) lies within the range in
\eqref{eq:rod-remainder-bound}, the remainder bound gives
\begin{equation}\label{eq:rod-app-deviation}
B\leq34(1-\lambda)^2 e^{-2L\sqrt{q(1-\lambda)}}.
\end{equation}
To ensure \(B\leq B_0\), it therefore suffices to take
\begin{align}
L\geq L_*(q,\lambda)
=\frac{\ln\big(1870\,m(\lambda)\big)}{2\sqrt{q(1-\lambda)}}.
\end{align}

We now estimate the error over the entire rod. Let \(w=y-y_\infty\),
\(x=2s\sqrt{q(1-\lambda)}\), and \(h(x)=\sech^2(x/2)\). At the center, \(w(0)=A-2(1-\lambda)\), so the turning-point relation
gives \(Aw(0)=2\lambda B\). Using this identity, we subtract the
infinite-rod equation \eqref{eq:rod-prototypical} from the finite-rod
equation \eqref{eq:rod-finite} to obtain
\begin{equation}\label{eq:errLN}
\mathcal{L}w=Bh+\frac{w(0)-B}{2}+\mathcal{N}(w),
\end{equation}
where the linear operator \(\mathcal{L}\) and the nonlinear term
\(\mathcal{N}(w)\) are given by
\begin{equation}
\mathcal{L}=\frac{d^2}{dx^2}-1+3h,
\qquad
\mathcal{N}(w)=\frac{w(0)^2+2Bw-3w^2}{4(1-\lambda)}.
\end{equation}

With  \(w'(0)=0\), variation of constants gives
\begin{equation}
w(x)=v(x)+\int_0^x G(x,t)\mathcal{N}(w(t))\,dt,
\end{equation}
where \(G\) is given by \eqref{eq:green-function} and the linear response is
\begin{equation}
v(x)=\frac{B-w(0)}{2}
+\frac{3w(0)-B}{2}\Big(h(x)+\frac{x}{2}h'(x)\Big).
\end{equation}
The end correction enters the estimate through \(v\) and the term
proportional to \(Bw\) in \(\mathcal{N}(w)\).

At the center, \(0\leq w(0)=2\lambda B/A\leq m(\lambda)B/(1-\lambda)\).
Suppose that
\begin{align}
|w(t)|\leq 1.05\,m(\lambda)B/(1-\lambda),\qquad  0\leq t\leq x.
\end{align}
Substituting the explicit linear response \(v\) into the
variation-of-constants formula and applying \eqref{eq:green-bound},
we obtain
\begin{align}
\frac{(1-\lambda)|w(x)|}{m(\lambda)B}
\leq 1+(1-h)\left(\frac{6}{110h}-\frac13\right)
\leq 1 \qquad \text{for } h\geq \frac16.
\end{align}
This closes the bootstrap in the region containing the center.

For \(h\leq 1/6\), we use \(y=2(1-\lambda)h+w\) to rewrite the
error equation \eqref{eq:errLN} in the form
\begin{align}
w''=a(x)w+f(x),
\end{align}
where the coefficient \(a(x)\) and the forcing term \(f(x)\) are
given by
\begin{align}
a(x) &= 1-\frac32\Big(h(x)+\frac{y(x)}{2(1-\lambda)}\Big)
+\frac{B}{2(1-\lambda)},\\
f(x) &= B\Big(h(x)-\frac12\Big)+\frac{\lambda B}{2(1-\lambda)}.
\end{align}
Here we used \(A=2(1-\lambda)+w(0)\) and \(Aw(0)=2\lambda B\)
to simplify \(f(x)\).

The bound \(y/[2(1-\lambda)]\leq 1/6+1/110\) at \(h=1/6\) holds
throughout the tail since \(y\) decreases toward the rod end.
Hence \(a>1/2-h\). At a positive interior maximum of \(w\),
\(aw\leq-f\leq B(1/2-h)\), so \(w\leq B\).
At a negative interior minimum, \(y/[2(1-\lambda)]\leq h\), giving
\(a\geq 1-3h\geq 1/2\). Thus \(a|w|\leq f\leq\lambda B/[2(1-\lambda)]\),
so \(|w|\leq\lambda B/(1-\lambda)\).
At the rod end, \(w\leq B\), and \(w'>0\) excludes a negative minimum
there. Combining these estimates with the center bound and using
\eqref{eq:rod-app-deviation} gives
\begin{align}
\sup_{0\leq s\leq L}|y(s)-y_\infty(s)|
\leq \frac{m(\lambda)B}{1-\lambda}
\leq 34(1-\lambda)m(\lambda)e^{-2L\sqrt{q(1-\lambda)}},\qquad L\geq L_*(q,\lambda).
\end{align}

\bibliographystyle{model5-names}
\bibliography{mybib}

\end{document}